\documentclass[11pt,a4paper,oneside]{article}

\usepackage{mymacro}
\usepackage{verbatim}
\usepackage{tikz}
\usetikzlibrary{arrows.meta}
 \usetikzlibrary{shapes.geometric}
\definecolor{myRed}{RGB}{150,22,22}
\protected\edef\mathcal{%
  \unexpanded\expandafter\expandafter\expandafter{%
    \csname mathcal \endcsname
  }%
}

\DeclareMathOperator{\dDisc}{dDisc}

\title{\textbf{Composite operators in $\boldsymbol{\CN=4}$ Super Yang-Mills}}

\author{Agnese Bissi$\,^{a,b,c}$,}
\author{Giulia Fardelli$\,^{d}$,}
\author{and Andrea Manenti$\,^{b}$,}

\affiliation{$^{a}$Abdus Salam International Centre for Theoretical Physics, \\
				Strada Costiera 11, 34151, Trieste, Italy}
\affiliation{$^{b}$Department of Physics and Astronomy, Uppsala University,\\
Box 516, SE-751 20 Uppsala, Sweden}
\affiliation{$^{c}$INFN, Sezione di Trieste, Via Valerio 2, I-34127 Trieste, Italy}

\affiliation{$^{d}$Department of Physics, Boston University, Boston, MA 02215, USA}

\makeatletter
\renewcommand{\@email}[1]{#1}
\makeatother

\abstract{
We consider four-point functions of protected, double- and single-trace operators in the large central charge limit. We use superconformal symmetry to disentangle the contribution of protected operators in the partial wave decomposition. With this information, we fix the non protected part of such correlators up to subleading order in the large central charge expansion. We particularly focus on the triple-trace sector of the correlator and comment on the connection to the holographic description of these correlators. }

\begin{document}

\maketitle

%\tableofcontents

\section{Introduction}

In recent years, significant progress has been made in the study of conformal field theories (CFTs), particularly through the development of the conformal bootstrap program~\cite{Rattazzi:2008pe,Poland:2018epd}. This approach leverages the symmetries of CFTs to impose strong constraints on their operator spectra and correlation functions. One of the most fruitful frameworks for applying these techniques is $\mathcal{N}=4$ Super Yang-Mills (SYM) theory with an SU$(N)$ gauge group, a  highly symmetric and integrable theory in four dimensions, which provides an ideal setting for the analytic conformal bootstrap~\cite{Bissi:2022mrs}.

A key focus within $\mathcal{N}=4$ SYM is the study of half-BPS operators, which belong to short multiplets of the superconformal group PSU$(2,2|4)$ and are annihilated by half of the supercharges. These operators form a protected subsector of the theory, with their conformal dimensions and three-point functions not acquiring quantum corrections. Most of the existing work has centred on analyzing four-point functions of single-trace scalar half-BPS operators, particularly in the regime of large central charge $c$, corresponding to the large rank of the gauge group $N$. In this regime, four-point functions of single-trace operators are related to the scattering amplitudes of single-particle string states in $AdS_5 \times S^5$ via the AdS/CFT correspondence.

The Operator Product Expansion (OPE) structure for these operators, along with constraints imposed by superconformal symmetry, allows the separation of contributions from short and semi-short protected operators and those that acquire anomalous dimensions, which belong to long multiplets~\cite{Dolan:2004iy}. In the large-$N$ expansion, the intermediate operators that appear in the conformal partial wave decomposition of unprotected four-point functions are multi-trace operators, corresponding to multi-particle states in the gravity dual setting.

Considerable progress has been made in systematically studying the anomalous dimensions of double-trace operators, which arise at leading order in the large-$N$ expansion and depend on the 't Hooft coupling $\lambda = g_{YM}^2 N$~\cite{Alday:2017xua,Bissi:2020wtv,Bissi:2020woe,Aprile:2017bgs,Aprile:2017xsp,Aprile:2017qoy,Aprile:2018efk,Aprile:2019rep,Drummond:2022dxw}. In the AdS/CFT correspondence, double-trace operators in the CFT side are dual to two-particle states in the AdS gravity theory. Understanding their dimensions and interactions provides insights into the dynamics of multi-particle states and the binding energies between particles in AdS space. Moreover, the study of double-trace operators is crucial in the attempt of reconstructing bulk interactions since the anomalous dimensions and OPE coefficients of double-trace operators contain information about the interactions in the bulk AdS theory. In addition they are extremely helpful in organizing and resumming contributions in the large-$N$ expansion, ensuring that the CFT data is consistent across different orders in $N^{-1}$. However, much less is known about the corrections to the dimensions of unprotected triple- and generically higher-trace operators. These operators are crucial to understand the structure of four-point functions, as they appear at order $N^{-6}$ and play a key role in determining the binding energy of multi-particle states in AdS.\footnote{Three-point functions involving triple-trace operators are present already at order $N^{-4}$, but their anomalous dimension appear only at next order, $N^{-6}$, when choosing a proper orthonormal basis.}

There are several ways to approach the study of higher-trace operators in the large $N$ expansion. One way is to study correlators of single-trace operators, with more than four fields.\footnote{Some works computing higher-point functions in $\CN=4$ SYM are~\cite{Goncalves:2019znr,Goncalves:2023oyx} and  in other contexts~\cite{Antunes:2021kmm,Harris:2024nmr,Kaviraj:2022wbw,Alday:2022lkk,Alday:2023kfm}.}In this setup there will be higher-trace operators already at subleading order, but the structure of the OPE and the proliferation of cross ratios make this approach quite impractical. Another way is to study four-point functions of half-BPS single-trace operators at order $N^{-6}$ in the large $N$ expansion~\cite{Bissi:2020wtv, Bissi:2020woe,Drummond:2022dxw}. This program has been carried out quite extensively, but, due to the fact that already at leading order in the large $N$ expansion one encounters a mixing problem~\cite{Alday:2017xua,Aprile:2017bgs,Aprile:2017xsp}, meaning that there is more than one operator with a definite set of quantum numbers, this approach turns out to be essentially hopeless.  Another way is to consider four-point functions of higher-trace operators, such that already at leading or subleading order there are multi-trace operators as intermediate states. This is one of the motivations of this paper and this is the setup that we are considering. 

Among multi-trace operators in $\mathcal{N}=4$ SYM,  there are several classes of them depending on how many supercharges they are annihilated by.  In a previous paper~\cite{Bissi:2021hjk}, we have considered the case of quarter-BPS operators, which are intrinsically double-trace operators. While there is a class of such operators which are scalars, the R-symmetry structure is much richer than the half-BPS counterpart but quite intricate, making the study of their four-point correlators cumbersome.  One of the main obstacles lies on the fact that superconformal blocks for quarter-BPS operators are not known.\footnote{For generic quarter-BPS operators, it is not even clear that it is possible to define superconformal blocks due to the fact that the three-point function of such operators is not protected.} However for large enough conformal dimension, there are also half-BPS operators which are double trace.  The first of such examples appears at dimension four.  If we denote schematically by $\phi$ the $\CN=4$ fundamental scalar,  at fixed $\Delta=4$, we can have two possible operators respecting all the symmetries: tr$\lsp \phi^4$ and tr$\lsp \phi^2$tr$\lsp\phi^2$.  The operators corresponding to eigenfunctions of the dilation generator can then  be constructed as  linear combinations of these building blocks with coefficients depending on $N$.  It turns out that a well-defined operator can be constructed entirely from the latter building block and it can be  interpreted as the dual of a two-particle state in AdS;   the same principle extends to operators with higher conformal dimensions. In this paper, we initiate the study of four-point functions involving this class of operators, starting with the simplest case: correlators including the double-trace operator of dimension four. We then use the power of superconformal symmetry together with the Lorentzian inversion formula to constrain the non-protected part of the correlators, providing information about the anomalous dimensions and three-point functions of higher-trace operators. 

The holographic counterpart of our computations involves bound states in the supergravity regime. As for the case of Kaluza-Klein modes of the graviton, we infer that  several Witten diagrams are protected and dictate the structure of the non protected ones. This perspective aligns closely with the recent results in~\cite{Aprile:2024lwy}, where the authors address the same problem on the gravity side using a more geometric approach. Their method constructs a Heavy-Heavy-Light-Light  correlator, from which one can access the four-point function of two dimension four double-trace operators and two dimension two single-particle operators up to $N^{-4}$.

The paper is organized as follows.  In Section~\ref{sec:setup}, we introduce the  dimension-four half-BPS operators we are going to study, the single-particle and the double-trace ones. Then we briefly outline our strategy to compute their correlator at leading order in the large charge expansion.  In Section~\ref{sec:twfour}, we compute the free-theory results necessary to extract OPE data of the exchanged  protected operators. We then resum their contributions to the correlators in Section~\ref{sec:resum}. By inputting this information in the double discontinuity, in Section~\ref{sec:dDisc}, we are able to compute the tree-level supergravity four-point functions consisting of two dimension-four operators and two stress-tensors.  In Section~\ref{sec:twist6}, we then extract new OPE data for twist-six long operators   and we find evidence for new triple-trace operators. 
We conclude in Section~\ref{sec:oneloop} with a discussion on how one could try to determine the correlator at higher order in $1/N$ and in Section~\ref{sec:futureDirections} with some future directions.  Several appendices contain technical details and explicit results.
\section{Set up}\label{sec:setup}
The goal of this paper is the study of four-point correlation functions of half-BPS operators of the same protected conformal dimension but different large-$N$ behavior.  In particular,  we would like to understand how to tackle the problem of dealing with double-trace operators, dual to multi-particle states, in the large $N$/large central charge limit.  We would like to analyze how their correlators differ from the ones involving  single-particle operators with the same dimension  and understand what is the appropriate basis of functions needed to describe them.

\subsection{Definition of the operators}
Half-BPS operators, which we will denote by $\CO_p(x_i, y_i)$, are scalar superprimaries  with protected dimension $\Delta=p$  transforming in the $[0,p,0]$ representation of the SU(4) R-symmetry.  The coordinates $x_i$ refer to the space-time insertion while $y_i$ are null vectors, encoding the R-symmetry polarizations.  On the field theory side, these operators can be built as traces, over the SU$(N)$ gauge group,  of various insertions of the real  scalars $\phi^I$, $I=1, \cdots, 6$, of $\CN=4$ SYM.  The first non-vanishing trace is simply given by tr$\lsp (y\cdot \phi)^2$, where we have contracted the SO(6) index $I$ with $y_I$.  Upon normalization, this corresponds to the $\CO_2$ operator,  superprimary of the stress-tensor multiplet. Similarly $\CO_3$ can be identified with the single trace operator built out of three $\phi$'s.  At $p=4$,  multiple ways of taking SU$(N)$ traces are possible,  in fact a dimension-four operator can either be built as $\tr (y\cdot \phi)^4$ or $(\tr(y\cdot \phi)^2)^2$.  In this case, the appropriate choice of operators which leads to an orthogonal and unit-normalized basis, is the following
\twoseqn{
\CO_4^{\mathrm{sp}}(x,y) &= \sqrt{\frac{4\lsp (N^2+1)}{(N^2-1)(N^2-4)(N^2-9)}}\left(t_4 - \frac{2\llsp N^2-3}{N(N^2+1)}\,t_2^2\right)\,,
}[]{
\CO_4^{\mathrm{dt}}(x,y) &= \sqrt{\frac{2}{N^4-1}}\,t_2^2\,.
}[][]
where
\eqn{
t_4 = \tr\, (y\cdot \phi)^4\,,\qquad t_2 = \tr\, (y\cdot \phi)^2\,.
}[]
It can actually be shown~\cite{Alday:2019nin,Aprile:2020uxk,Aprile:2019rep,Aprile:2018efk} that only the first linear combination is dual to a proper single-particle state in AdS, from which the superscript ``sp'' arises.  In contrast,  the other operator corresponds to a pure double-trace state, which is why we labeled  it ``dt''.

By construction we have
\eqn{
\langle \CO_4^{\mathrm{sp}}\CO_4^{\mathrm{sp}}\rangle = \frac{y_{12}^4}{(x_{12}^2)^4}\,,\qquad
\langle \CO_4^{\mathrm{dt}}\CO_4^{\mathrm{dt}}\rangle = \frac{y_{12}^4}{(x_{12}^2)^4}\,,\qquad
\langle \CO_4^{\mathrm{dt}}\CO_4^{\mathrm{sp}}\rangle = 0\,.
}[]
and three-point functions
\eqn{
\langle \CO_2\CO_2\CO_4^{\mathrm{sp}}\rangle = 0\,,\qquad
\langle \CO_2\CO_2\CO_4^{\mathrm{dt}}\rangle = \frac{\sqrt{2}(N^2+1)}{\sqrt{N^4-1}}\frac{y_{13}^2y_{23}^2}{(x_{13}^2 x_{23}^2)^2}\,.
}[]
Notice that only $\CO_4^{\rm dt}$ is exchanged in the OPE of $\CO_2\times\CO_2$~\cite{Dolan:2004iy}.
These are the operators appearing in the four-point functions that we will consider in this paper, together with  single trace $\CO_2$. In Appendix \ref{highertrace} we construct the basis for operators of $p \geq 4$ and claim that there always exists at least one well defined operator made of double-trace operators only. 
\subsection{Strategy}
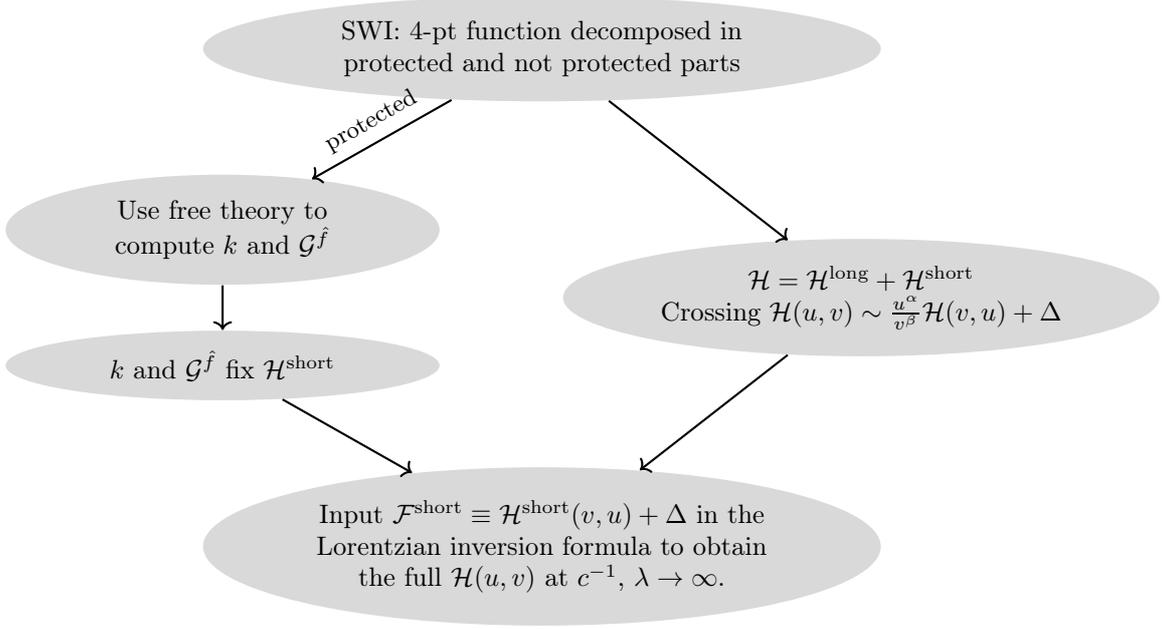
\begin{figure}[h!] \centering
\begin{tikzpicture}
\node [ellipse, fill=lightgray!60!white] (SCFT) at  (0,0.8) {\begin{minipage}{0.4\textwidth} {\small \begin{center}
SWI: $4$-pt function  decomposed in protected and not protected parts
\end{center}}
\end{minipage}
};
\node [ellipse, fill=lightgray!60!white] (FS) at (-4.2,  -1.6){\begin{minipage}{0.25\textwidth} {\small \begin{center}
Use free theory to compute $k$ and $\CG^{\hat{f}}$
\end{center}}
\end{minipage}
};
\node [ellipse, fill=lightgray!60!white] (short) at (-4.2,  -3.4){\begin{minipage}{0.25\textwidth} {\small \begin{center}
$k$ and $\CG^{\hat{f}}$  fix $\CH^{\rm short}$
\end{center}}
\end{minipage}
};
\draw[thick,->] (FS)--(short);
\node [ellipse, fill=lightgray!60!white] (CB) at (4.2,  -2.5){\begin{minipage}{0.35\textwidth} {\small \begin{center}
$\CH=\CH^{\rm long}+\CH^{\rm short}$\\
Crossing $\CH(u,v)\sim \frac{u^\alpha}{v^\beta}\CH(v, u)+\Delta$
\end{center}}
\end{minipage}
};
\node [ellipse, fill=lightgray!60!white] (QG) at  (0,-5.8) {\begin{minipage}{0.4\textwidth} {\small \begin{center}
Input $\CF^{\rm short}\equiv\CH^{\rm short}(v,u)+\Delta$ in the Lorentzian inversion formula to obtain the full $\CH(u,v)$ at $c^{-1}$, $\lambda\to\infty$.
\end{center}}
\end{minipage}
};
\draw[thick,->] (SCFT)--(FS) node[pos=0.5,sloped,above] {\footnotesize protected};
\draw[thick,->] (SCFT)--(CB);
\draw[thick,->] (short)--(QG);
\draw[thick,->] (CB)--(QG);
\end{tikzpicture}\caption{A schematic representation of the strategy will be using to construct correlators at large $c$ and large $\lambda$. }\label{fig:scheme}
\end{figure}
In this paper we discuss  the large central charge $c=\frac{N^2-1}{4}$ behavior of correlators involving  half-BPS, dimension-four, operators $\CO_4^{\mathrm{sp}}$ and $\CO_4^{\mathrm{dt}}$. We will denote all the quantities relative to the first/last one with a superscript ``sp''/``dt''.\\ Inspired by the work in~\cite{Alday:2017vkk}, in the case of  half-BPS operators of dimension two, we reconstruct the tree-level correlator in the supergravity limit, i.e. at leading order at large $c/N$ and t' Hooft coupling $\lambda\to \infty$,  entirely from the knowledge of the protected subsector of the four-point function. 
In the case of four identical $\CO_2$'s,  it was shown that short and semishort operators exchanged in the OPE are  the only ones producing  singular, therefore non-vanishing,  contributions inside the Lorentzian inversion formula~\cite{Caron-Huot:2017vep} and therefore they can be used  to fix the entire correlator.

In the chart in Fig.~\ref{fig:scheme}, we have summarized how we are going to extend these results in the case of $\langle \CO_2\CO_2 \CO_4^{\mathrm{sp}}\CO_4^{\mathrm{sp}}\rangle$ and $\langle \CO_2\CO_2 \CO_4^{\mathrm{dt}}\CO_4^{\mathrm{dt}}\rangle$.  Consider a four-point function of generic half-BPS operators
\eqn{
\langle \CO_{p_1}(x_1,y_1)\CO_{p_2}(x_2,y_2)\CO_{p_3}(x_3,y_3)\CO_{p_4}(x_4,y_4)\rangle = \CK_{\{p_i\}}(x_i,y_i) \;\CG_{\{p_i\}}(z,\zb;\alpha,\alphab)\,,
}[]
where $\CK$ is a kinematic prefactor and $\CG$ is a function of conformal and SU$(4)$ R-symmetry cross ratios --- see Appendix~\ref{app:conv} for more details and conventions.  Superconformal Ward Identities (SWI) allow us to further decompose $\CG$ to make explicit the contributions from  short, semi-short and long superconformal multiplets exchanged in the OPE~\cite{Dolan:2004iy}
\eqna{
(z \zb)^{\frac{p_3-p_4}{2}}\CG_{\{p_i\}}(z,\zb;\alpha,\alphab) &= k_{\{p_i\}} + \CG_{\{p_i\}}^{\hat{f}} + R_{\{p_i\}}\CH_{\{p_i\}}(z,\zb;\alpha,\alphab)\,.
}[WIsplit]
The functions $k$ and $\CG^{\hat{f}}$, defined in~\eqref{kandF} and~\eqref{otherkandF}, contain only short and semi-short multiplets. Since these operators do not depend on the coupling $\lambda$, these functions are fixed from free field results.  Long supermultiplets can only appear inside $\CH$, which, however, can still receive protected contributions. We will denote this part of the reduced correlator by $\CH^{\rm short}$ and we will show how to determine it from the knowledge of $k$ and $\CG^{\hat{f}}$. So the full reduced four-point function can be decomposed as
\eqn{
\CH_{\{p_i\}}(z,\zb;\alpha,\alphab)  = \CH^{\mathrm{short}}_{\{p_i\}}(z,\zb;\alpha,\alphab)  + \CH^{\mathrm{long}}_{\{p_i\}}(z,\zb;\alpha,\alphab) \,,
}[]
where $\CH^{\mathrm{long}}_{\{p_i\}}(z,\zb)$ has contributions only from long multiplets and the three-point functions are real and positive in a reflection-positive configuration. Furthermore, at large central charge and in the supergravity limit, there are no single-trace operators in $\CH^{\rm long}$ and the twist gap is $\min(p_1+p_2,\,p_3+p_4)$.  Finally, due to the decomposition in~\eqref{WIsplit}, $\CH$ is no longer crossing symmetric by itself. 
Let us now assume that there is only one SU(4) representation exchanged in $\CH$, as it is in the case  we study,  so that we can suppress the R-symmetry dependence, then crossing reads
\eqna{
\CH^{\mathrm{long}}_{p_1p_2p_3p_4}(z,\zb) +\CH^{\mathrm{short}}_{p_1p_2p_3p_4}(z,\zb) &= \frac{u^{\frac12(p_1+p_2)}}{v^{\frac12(p_2+p_3)}} \CH^{\mathrm{short}}_{p_3p_2p_1p_4}(1-z,1-\zb) + \Delta_{p_1p_2p_3p_4}\\ &\;\quad
+\frac{u^{\frac12(p_1+p_2)}}{v^{\frac12(p_2+p_3)}} \CH^{\mathrm{long}}_{p_3p_2p_1p_4}(1-z,1-\zb)\,,
}[HcrossRest]
where $\Delta_{p_1p_2p_3p_4}$ can be obtained by crossing the full four-point function as usual, knowing  $k$ and $\CG^{\hat{f}}$.
Similar to the discussion in~\cite{Alday:2017vkk},  in the following we will use 
\eqn{
\CF_{p_1p_2p_3p_4}^{\mathrm{short}}(z,\zb) \equiv s \frac{u^{\frac12(p_1+p_2)}}{v^{\frac12(p_2+p_3)}} \CH^{\mathrm{short}}_{p_3p_2p_1p_4}(1-z,1-\zb) + \Delta_{p_1p_2p_3p_4}(z,\zb)\,.
}[FshortDef]
inside the Lorentzian inversion formula to reconstruct the full tree-level supergravity correlator. With respect to the four  $\CO_2$'s correlator,  the cases under consideration are complicated by the fact that crossing mixes two channels, i.e.  four-point function where the same operators appear in different orderings, thus making necessary  to resum four sets of infinitely many protected operators, two for each channel.  Moreover, the exchange of potentially new protected twist-four operators inside $\CH_{4224}^{\rm short}$ presents an additional technical challenge, that we will be able to circumvent as explained in the following section.

\section{Free theory}\label{sec:twfour}
As explained before to construct $k$ and $\CG^{\hat{f}}$ we only need free fields results.  So we can study the free theory part of the correlators, and then  use superconformal Ward identities to identify which contributions  come from the protected operators.  The correlator in the free theory is simply computed by taking  Wick contractions
\eqn{
\contraction[1.4ex]{y_1\cdot}{\phi}{{}^A(x_1)\; y_2\cdot{}}{\phi}y_1\cdot\phi^A(x_1)\; y_2\cdot\phi^B(x_2) = \frac{y_{12}}{x_{12}^2}\,\delta^{AB}\,,\qquad A,B=1,\ldots,N^2-1\,.
}[]
To avoid clattering we collect all the explicit results in Appendix~\ref{app:free}. While we obtain explicit expressions for the function $\mathcal{H}^{\text{sp/dt}}$ in the free theory --- see~\eqref{chfree2244sp},  \eqref{chfree2244dt}, \eqref{chfree4224sp}, \eqref{chfree4224dt} --- we would still need to find three-point functions from which we will be able to isolate  $\mathcal{H}^{\rm short \text{,sp/dt}}$ by imposing the correct expansion in terms of superconformal blocks with positive OPE coefficients.  Moreover, as anticipated before, additional protected contributions can appear in $\CH_{4224}^{\rm short}$. To address this problem, we will  extract the relevant OPE data directly by analyzing $\langle \CO_2\CO_2\CO_2\CO_4\rangle$.
This correlator gives us information about the protected subsector because it falls into the  category of ``next-to-extremal'' correlator,  i.e. four-point functions in which the difference between the biggest external dimension and the remaining ones is equal to -2. This class of correlators  has been shown~\cite{Erdmenger:1999pz,Eden:2000gg} to be non-renormalised:  their expressions are fixed by free-theory results, they do not receive any quantum corrections and all the exchanged operators are  protected.  Morever, this example will  serve as a good warm-up to illustrate how SWI operates and how information about protected subsectors can be  extracted in a simpler setup.

\subsection{Next-to-extremal correlator} \label{sec:NTE}
The four-point function involving three $\CO_2$ and one $\CO_4$ can be written as
\eqna{
\langle \CO_2  \CO_2  \CO_2 \CO_4 \rangle &= \CK_{2224}(x_i, y_i)\CG_{2224}(z, \zb, \alpha, \alphab)\, ,\\
\CG_{2224}(z, \zb, \alpha, \alphab)&=\sum_{n=0}^1\sum_{m=0}^n \sum_{\Delta, \ell} \lambda_{22\CO} \lambda_{\CO 24}\lsp  Y_{n,m}^{(0, -2)} g_{\Delta, \ell}^{(0, -2)}(z, \zb)\, ,
}[]
where we have denoted with $\lambda_{ij\CO}$ the OPE coefficient between $\CO_i, \, \CO_j$ and the exchanged operator $\CO_{\Delta, \ell}$, which can transform in the  $[0,2,0], \, [1,2,1]$ or $[0,4,0]$ SU(4) representations. Using Wick contractions it is straightforward to compute
\eqna{
\langle\CO_2\CO_2\CO_2\CO_4^{\mathrm{sp}} \rangle&=0\,, \\
\CG_{2224}^{\mathrm{dt}}(z, \zb, \alpha, \alphab)&=
\frac{\sqrt{4 c+2}}{c}\left(u^2\lsp   \alpha \alphab +\frac{ u^2}{v}(\alpha -1) (\alphab-1)+u  \right)\, .
}[]
The vanishing of the  four-point function with the single-particle operator  has already been observed in~\cite{Aprile:2020uxk}, where the authors can more generally prove that any ``near-to-extremal'' correlator, involving half-BPS operators and a single-particle operator,  vanishes provided that the single-particle operator has the largest charge. 
Moving on to the correlator involving $\CO_4^{\rm dt}$, we can use the expansion in~\eqref{WIsplit} to determine
\threeseqn{
k^{\mathrm{dt}}_{2224}&= \frac{3\sqrt{4 c+2}}{c}\, ,
}[]
{
\hat{f}^{\mathrm{dt}}_{2224}&=\frac{\sqrt{4 c+2}}{c}\lsp \frac{z(z-2)}{z-1}\, ,
}[]
{
\CH^{\mathrm{dt}}_{2224}&=0\, .
}[][]
Then we need to expand  $k$ and $\hat{f}$ in superconformal blocks as explained in Appendix~\ref{app:conv}. One may wonder why this is necessary and it is not enough to simply expand the free $\CG_{2224}^{\rm dt}$ in blocks. The reason is that only by using the expansion satisfying SWI,  we are guaranteed to extract the OPE coefficients of superconfomal primaries.  $\hat{f}^{\mathrm{dt}}_{2224}$ can be expanded in one-variable blocks as in~\eqref{fexp}
\eqna{
\hat{f}^{\mathrm{dt}}_{2224}&=\sum_{\ell=0}^{\infty} b_{\ell} \lsp z^{-1} k_{\ell+2}^{(0, -2)}(z)\,,\\
b_\ell& =\frac{\sqrt{4 c+2}}{c} \left(1+(-1)^\ell\right) \frac{(\ell +1)! (\ell +2)!}{(2 \ell +2)!}\, .
}[]
If we plug this expansion back to $\CG^{\hat{f}}$, we can isolate contributions from the different R-symmetry channels. Focusing on the $[0,2,0]$,  which is the one relevant for the following discussion, if we denote by $\CG_{2224,00}$ this contribution to the full four-point function, we find
\eqna{
z \zb \,  \CG^{\hat{f}, \lsp {\mathrm{dt}}}_{2224,00}&=b_0\left(- g_{2,0}^{(0, -2)}+\frac{2}{175} g_{6,2}^{(0, -2)}\right)+\sum_{\ell=0}^\infty (-2)^\ell b_{\ell+2} \Bigg(g_{\ell+4, \ell}^{(0, -2)} +\\&\quad  \, \frac{6(\ell+2)(\ell+5)}{5(2\ell+5)(2\ell+9)}g_{\ell+6, \ell+2}^{(0, -2)}+\frac{(\ell +3) (\ell +4) (\ell +5) (\ell +6)}{(2 \ell +7) (2 \ell +9)^2 (2 \ell +11)}g_{\ell+8, \ell+4}^{(0, -2)}\Bigg)\, .
}[Gf2224]
We can now read off the OPE coefficients of the various protected operators contributing.  
The only twist two term combines with the one coming from the $k^{\mathrm{dt}}_{2224}$-piece and together they give the OPE coefficient corresponding to the exchange of $\CO_2$ itself, which is the only protected operator at twist 2 in the $[0,2,0]$\footnote{This result is consistent with the explicit computation of the three-point functions
\eqna{
\langle \CO_2 \CO_2 \CO_2 \rangle&=\sqrt{\frac{2}{c}} \frac{y_{12}y_{13}y_{23}}{x_{12}^2x_{13}^2x_{23}^2}\, ,\\
\langle \CO_2 \CO_2 \CO_4^{\mathrm{dt}} \rangle&= \sqrt{\frac{2c+1}{c}} \frac{y_{13}^2 y_{23}^2}{(x_{13}^2x_{23}^2)^2}\, .
}[]}
\eqna{
\lambda_{222}\lambda_{224^{\mathrm{dt}}}=\frac{\sqrt{4c+2}}{c}\, .
}[]
Finally, from the expression above we can  extract the OPE coefficient involving the twist-four operators we are after, which we can identify with  $\CC_{[0,2,0], \ell}$~\cite{Dolan:2002zh} 
\eqna{
\lambda_{22\CC}\lambda_{\CC 24}=(-2)^\ell b_{\ell+2}\,.
}[lambdaC24]
The only thing left is to extract $\lambda_{\CC 24}$ by diving by the OPE coefficient $\langle \CO_2 \CO_2 \CC_{[0,2,0], \ell} \rangle$, which can be taken for example from~\cite{Dolan:2004iy}. By properly taking into account normalization of the SU(4) tensor structure --- see Appendix~\ref{app:tensors} --- we obtain:
\eqna{
 \lambda_{\CC 24^{\mathrm{dt}}}^2=\left(\frac{(-2)^{\ell+2} b_{\ell+2}}{\sqrt{\lambda_{22\CC}^2}}\right)^2 \xrightarrow{\text{ large } c\,} 2^{\ell } \left((-1)^{\ell }+1\right)\frac{(\ell +2)! (\ell +4)!}{c \lsp (2 \ell +6)!}.
}[lC24dtSqr]
Notice that, given how we obtained it,  this result applies only to even spin. Luckily, when $\ell$ is odd these operators do not appear, so the OPE coefficients in~\eqref{lC24dtSqr} are the only ones we will need to fix $\CH_{4224}^{\rm short, dt}$ in Sec.~\ref{Sec:channel4224}.

\section{Short contributions to \texorpdfstring{$\boldsymbol{\CH}$}{H} and resummation}\label{sec:resum}
The Ward identity splitting of~\eqref{WIsplit} commutes nicely with the super-Casimir of ${\CN=4}$ which means that all contributions from long multiplets will appear inside $\CH_{\{p_i\}}$ only. The converse is not true however, the reduced correlator $\CH_{\{p_i\}}$ will also receive some contributions from short multiplets that we need to isolate. This is one way to see this concretely: a single protected contribution inside $\hat{f}_{\{p_i\}}$ takes the form
\eqn{
\hat{f}_{\{p_i\}}(z)\big|_{n,\ell} = b_{\{p_i\}}(n,\ell)\,y_n^{(p_{12},p_{34})}(\alpha)\,\mathfrak{g}_\ell^{(p_{12},p_{34})}(z)\,.
}[]
When we plug this inside $\CG^{\hat{f}}_{\{p_i\}}$ in~\eqref{WIsplit} and try to expand the resulting four-point function in conformal blocks as~\eqref{allCBexp} we run into problems. We may either find operators that are below or at unitarity, or we could even obtain a function that cannot be expanded in conformal blocks at all. The reason is that the protected multiplet that gives the $n,\ell$ contribution to $\hat{f}$ also contributes to $\CH_{\{p_i\}}$
\eqn{
\CH_{\{p_i\}}(z,\zb)\big|_{n,\ell} = A_{\{p_i\}}(n,\ell)\,Y_{n',m'}^{(p_{12},p_{34})}(\alpha,\alphab)\,G_{\ell+\tau,\ell}^{(p_{12},p_{34})}(z,\zb)\,,
}[]
for some coefficient $A_{\{p_i\}}(n,\ell)$ and some appropriate numbers $n',m'$ and $\tau$ which we do not specify in this general discussion. Plugging both $\hat{f}_{\{p_i\}}$ and $\CH_{\{p_i\}}$ inside~\eqref{WIsplit} leads to a correct conformal block expansion.
We denote as $\CH^{\mathrm{short}}_{\{p_i\}}$ the collection of all contributions from short operators
\eqn{
\CH^{\mathrm{short}}_{\{p_i\}}(z,\zb;\alpha,\alphab) = \sum_{n',m',\tau} A_{\{p_i\}}(n,\ell)\,Y_{n',m'}^{(p_{12},p_{34})}(\alpha,\alphab)\,G_{\ell+\tau,\ell}^{(p_{12},p_{34})}(z,\zb)\,.
}[]
In this section we will focus in identifying this part of the correlator in each of the two channels by resumming the contributions from the protected operators. 
\subsection{Channel $\langle 2244\rangle$}
Following the conventions in Appendix~\ref{app:conv},  $\hat{f}$ can be expanded in one-variable conformal and R-symmetry blocks as
\eqna{
\hat{f}_{2244}(z)=\sum_{n=0}^1 \sum_{\ell=0}^\infty b_{2244}(n, \ell)y_n^{(0,0)} \mathfrak{g}_{\ell}^{(0,0)}(z)\, ,
}[]
where the explicit expressions for $b_{2244}$ for single-particle and double-trace operators are recorded in~\eqref{b2244sp} and~\eqref{b2244dt}.
In this correlator, a contribution to the $n=1$ sector of $\hat{f}_{2244}$ yields operators of twist zero in the fully expanded $\CG_{2244}$, while a contribution to the $n=0$ sector yields operators of twist two. In general, while a long multiplet only contributes to $\CH$, a protected multiplet will contribute to $k$, $\hat{f}$ and $\CH$ simultaneously. The precise relation between all such contributions can be fixed by solving the superconformal Casimir equation, but we have a shortcut. Indeed, we also know that, at the end of the day, any given multiplet cannot yield twist-zero or twist-two operators when expanded in the full $\CG$ for the interacting theory~\cite{Maldacena:2011jn,Alba:2013yda}. This criterion allows us to fix the contribution to $\CH_{2244}$ corresponding to any single contribution to $\hat{f}_{2244}$.\\
Imposing twist zero cancellation we have, for any positive odd $\ell$,\footnote{Multiplets $\CC_{[p,q,p]\ell}$ for negative $\ell$ are a slight abuse of notation. They can be written in terms or sums of other shorter multiplets as done in~\cite{Dolan:2004iy}.}
\eqn{
\CG(\CC_{[0,2,0]\ell-3}) = \left\lbrace\begin{aligned}
\CH_{2244} &\to b_{2244}(1,\ell)\, (-2)^{\ell-1} \lsp G_{\ell-1,\ell-1}^{(0,0)}(z,\zb)\,,
\\\hat{f}_{2244} &\to b_{2244}(1,\ell)\left(\alpha-\frac12\right)\,k_{\ell+1}^{(0,0)}(z)\,.
\end{aligned}\right.
}[]
While imposing cancellation of twist two, for any integer spin $\ell \geq2$, leads to
\eqn{
\CG(\CC_{[1,0,1]\ell-2}) = \left\lbrace\begin{aligned}
\CH_{2244} &\to b_{2244}(0,\ell)\,(-1)(-2)^{\ell-2} \, G_{\ell,\ell-2}^{(0,0)}(z,\zb)\,,
\\\hat{f}_{2244} &\to b_{2244}(0,\ell)\,k_{\ell+1}^{(0,0)}(z)\,.
\end{aligned}\right.
}[]
To reiterate the reasoning of before: if we give an additional contribution to $\CH_{2244}$ as follows
\eqn{
\CH_{2244} \to a_{2244}(\ell+2,\ell) \, G_{\ell+2,\ell}^{(0,0)}(z,\zb)\,,
}[]
we are modifying the OPE coefficient of the multiplet $\CC_{[1,0,1]\ell}$ but \emph{also} generating a $\CC_{[0,0,0]\ell}$ multiplet with coefficient $a_{2244}(\ell+2,\ell)$. Fixing the twist two OPE coefficient to be just $-(-2)^\ell b_{2244}(0,\ell+2)$ is equivalent to requiring the cancellation of the $\CC_{[0,0,0]\ell}$ multiplet.

In principle, in the formulas above one should compute also the contribution to $k_{2244}$, but for our purposes we are interested in $\CH_{2244}$ only so that is not important.

All in all, the final result that we need is the following
\eqn{
\CH_{2244}^{\mathrm{short}}= \sum_{\ell=0,2,4,\ldots} (-2)^\ell\left( b_{2244}(1,\ell+1)\, G^{(0,0)}_{\ell,\ell} - b_{2244}(0,\ell+2)\,G^{(0,0)}_{\ell+2,\ell}\right)\,.
}[]
After performing the sum over $\ell$ we obtain a close form expression for both $\CH_{2244}^{\mathrm{short,\,sp}}$ and $\CH_{2244}^{\mathrm{short,\,dt}}$.  The explicit results are collected in Appendix~\ref{app:Hshort}.

\subsection{Channel $\langle 4224\rangle$} \label{Sec:channel4224}
Here the situation is different with respect to the case with $p_{34}=0$. The first difference is that the contributions to $\hat{f}$ do not give twist zero and twist two but rather give twist two and twist four. The second difference is that the twist-two contributions are actually ill defined, in the sense that they give rise to blocks in $\CG_{4224}$ which are divergent. This last point is nothing to be worried about: we know that such contributions must be canceled in the end. The only inconvenience is that we cannot impose the cancellation by expanding in conformal blocks but we rather have to expand in powers of $u$ and cancel all terms linear in $u$. The presence of twist-four operators, on the other hand, presents an additional issue. The problem is that the superconformal block expansion of the free theory is ambiguous: appropriately fine-tuned sums of protected blocks of different spin can mimic a long block at unitarity threshold. This phenomenon is usually termed ``multiplet recombination.'' The naive expectation is that every contribution that can be repackaged as a long multiplet always acquires an anomalous dimension in the interacting theory and therefore every multiplet recombination that can happen will actually happen. It was observed in~\cite{Doobary:2015gia} that this is not always true. Due to this problem we cannot trust the free theory to give us the correct OPE coefficients for the protected multiplets. Therefore we resort to four-point functions which are entirely protected, namely $\langle\CO_2\CO_2\CO_2\CO_4^{\mathrm{dt}}\rangle$ and $\langle\CO_2\CO_2\CO_2\CO_4^{\mathrm{sp}}\rangle$,  introduced in Sec.~\ref{sec:twfour}.

Just as we did before, we impose the cancellation of twist two and we obtain
\eqn{
\CG(\CC_{[0,4,0]\ell-3}) = \left\lbrace\begin{aligned}
\CH_{4224} &\to b_{4224}(1,\ell)\, (-2)^{\ell-1} \lsp G_{\ell+1,\ell-1}^{(2,-2)}(z,\zb)\,,
\\\hat{f}_{4224} &\to b_{4224}(1,\ell)\left(\alpha-\frac14\right)\,k_{\ell+2}^{(2,-2)}(z)\,.
\end{aligned}\right.
}[]
Unlike the previous case, we cannot declare that all twist-four operators must disappear. It is however useful to know the contributions to $\CH_{4224}$ that cancel exactly those from $\hat{f}_{4224}$ so that we will have an easier time adding the correct one we computed earlier. The result for all $\ell\geq 2$ is
\eqn{
\CG(\CC_{[1,2,1]\ell-2}) = \left\lbrace\begin{aligned}
\CH_{4224} &\to b_{4224}(0,\ell)\,(-1)(-2)^{\ell-2} \, G_{\ell+2,\ell-2}^{(2,-2)}(z,\zb)\,,
\\\hat{f}_{4224} &\to b_{4224}(0,\ell)\,k_{\ell+2}^{(2,-2)}(z)\,.
\end{aligned}\right.
}[]
Note that this only cancels the tower of higher-spin twist-four operators but leaves nonzero the contribution of the $[0,4,0]$ half-BPS operator itself to the OPE.
Let us also repeat the same comment that we did in the previous subsection: modifying the twist-four coefficient as
\eqn{
\CH_{4224} \to a_{4224}(\ell+4,\ell) \, G_{\ell+4,\ell}^{(2,-2)}(z,\zb)\,,
}[]
will modify the $\CC_{[1,2,1]\ell}$ contribution as well as generating a $\CC_{[0,2,0]\ell}$ new multiplet. Fixing the coefficient at twist four to be exactly $-(-2)^\ell b_{4224}(0,\ell+2)$ is equivalent to set the $\CC_{[0,2,0]\ell}$ multiplets to zero.

Finally, we can define $\CH^{\mathrm{short}}_{4224}$ by summing up all these contributions. Let us define $\hat\CH^{\mathrm{short}}_{4224}$ to be the short reduced correlator assuming the absence of $\CC_{[0,2,0]\ell}$ and $\CH^{\mathrm{short}}_{4224}$ to be the full one.  Let us also call $\delta b_{4224}(\ell)$ the squared OPE coefficient between $\CO_2$, $\CO_4$ and the superprimary of $\CC_{[0,2,0]\ell}$, which has dimension $4+\ell$ and transforms in the $[0,2,0]$.
With these definitions we have
\twoseqn{
\hat\CH_{4224}^{\mathrm{short}}&= \sum_{\ell=0}^\infty (-2)^\ell\left( b_{4224}(1,\ell+1)\, G^{(2,-2)}_{\ell+2,\ell} - b_{4224}(0,\ell+2)\,G^{(2,-2)}_{\ell+4,\ell}\right)\,,
}[]{
\CH_{4224}^{\mathrm{short}} &= \hat\CH_{4224}^{\mathrm{short}} + \sum_{\ell=0}^\infty \delta b_{4224}(\ell)\, G_{\ell+4,\ell}^{(2,-2)}\,.
}[CHnothat][]
The expressions are too lengthy, so we report them explicitly in Appendix~\ref{app:Hshort}. 
For the additional twist-four contribution, in Sec.~\ref{sec:NTE}, we have shown that for the single-particle case, no other terms need to be added. In fact $\delta b_{4224}^{\rm sp}=0$, so that $ \hat\CH_{4224}^{\mathrm{short, sp}} = \CH_{4224}^{\mathrm{short, sp}} $. On the contrary,  for the correlator involving the double-trace operators, also $\CC_{[0,2,0]\ell}$ are exchanged. If we resum them as prescribed by~\eqref{CHnothat} with the  OPE coefficient computed in~\eqref{lC24dtSqr}, we get
\eqna{
\delta\CH_{4224}^{\mathrm{short, dt}}|_{\CC_{[0,2,0]\ell}}&=\frac{-40}{3\lsp  c \lsp z^3 \zb^2 (z-\zb)}  \left( \zb \left(\zb^2-9 \zb+6\right)+\left(5 \zb^2-12 \zb+6\right) \log (1-\zb) \right)\times\\ 
&\quad \,\left(z \left(-17 z^2+42 z-24\right)+6 (z-4) (z-1)^2 \log (1-z) \right)+ z \leftrightarrow \zb\, .
}[H4224shortContribExpl] 

\section{Double discontinuity}\label{sec:dDisc}
Over the last few years, the Lorentzian inversion formula~\cite{Caron-Huot:2017vep} has been proven to be a very powerful tool to determine new OPE coefficients and ``bootstrap'' correlators in a variety of contexts. The main advantage comes from the fact that, in this setup, the OPE data are encoded in a function expressed in terms of the double discontinuity (dDisc) of the correlator. Being affected just by the singular behavior of the correlator as $\zb\to 1$, dDisc does not depend on the knowledge of the full four-point function but rather on a subsector of exchanged operators, which is usually easier to compute. This has been particularly advantageous when applied to supersymmetric theories at strong coupling and at large central charge. Here, it has been shown how the four-point function at order $c^{-k}$ can be possibly reconstructed from its dDisc, which is fixed by quantities computed at previous orders, through a mechanism similar to unitarity cuts and dispersion relation in the amplitudes context~\cite{Aharony:2016dwx, Alday:2017vkk,Bissi:2020woe,Bissi:2020wtv}. In particular in~\cite{Alday:2017vkk},  the full tree-level supergravity correlator $\CH_{2222}$ in $\CN=4$ SYM was reconstructed from its dDisc, which was shown to arise entirely from the exchange of protected operators in the OPE.  Following the same philosophy ---  that the protected subsector can determine the full supergravity correlator ---  in this section we analyze the dDisc of the four-point functions involving both single-particle and double-trace operators.  Compared to the original work,  this case is complicated by the presence of  pairwise identical operators,  requiring different channels to be studied separately.  The correlator $\langle \CO_2\CO_2\CO_4\CO_4\rangle$, being symmetric under $1\leftrightarrow 2$ exchange, is comparatively simpler, so we will focus on this channel in the following discussion. For all technical details, we refer to Appendix~\ref{app:inversion}. Schematically, the OPE density computed by the Lorentzian inversion formula in this channel can be expressed as:
\eqna{
c_{2244}\sim (1+(-1)^\ell) \int \text{dDisc}\left[  \frac{u^2}{v^3}\CH_{4224}(v, u)+\Delta_{2244}\right]\, ,
}[]
where we have used the fact that $\CH$ is not perfectly crossing symmetric as in~\eqref{HcrossRest}.  In this case and  up to order $1/c$,  the only non-vanishing contributions to dDisc can come from negative powers of $v$.  So in particular they can only come,  upon crossing, from the exchange of twist-two and four operators  inside $\CH_{4224}$. But these are exactly the protected contributions we have computed in the previous section. Therefore
\eqna{
c_{2244}\sim (1+(-1)^\ell) \int \text{dDisc}\left[ \CF^{\rm short}_{2244}\right]\, ,
}[cFshort]
where $\CF^{\rm short}$ is defined in~\eqref{FshortDef}.  In the cases of interests, it is easy to evaluate dDisc from the expressions in~\eqref{CHshort4224sp} and~\eqref{CHshort4224dt}.  We will keep unevaluated the discontinuities
\eqna{
\dDisc\lsp\left[\left(\frac{\zb}{1-\zb}\right)^n \zb^{\lsp-\frac{p_{34}}2}\right]\,,}[]
and denoting
\eqn{
\zeta = \frac{z}{1-z}\,,
}[]
the results are
\twoseqn{
\dDisc\left[\hat\CF_{2244}^{\mathrm{short,\,sp}}\right] &= \frac1c\, \dDisc\left[\frac{\zb}{1-\zb}\right]\left(
6\zeta-8\zeta^2+12\zeta^3-24\zeta^4-24\zeta^5 \log z
\right)\,,
}[]{
\dDisc\left[\hat\CF_{2244}^{\mathrm{short,\,dt}}\right]  &=\left(2+\frac1c\right)\left(\zeta \dDisc\left[\frac{\zb^2}{(1-\zb)^2}\right] - \zeta^2 \dDisc\left[\frac{\zb}{1-\zb}\right]\right)\\&\;\quad
 + \frac1c\dDisc\left[\frac{\zb}{1-\zb}\right]\left(
6\zeta-8\zeta^2+12\zeta^3-24\zeta^4-24\zeta^5\log z
\right)\,.\nonumber
}[][]
The contribution coming from~\eqref{H4224shortContribExpl} gives instead
\eqna{
\dDisc \left[ \delta \CF_{2244}^{\mathrm{short,\,dt}}|_{\CC_{[0,2,0], \ell}} \right]&=\frac{4}{c}\dDisc\left[\frac{\zb}{1-\zb}\right](z-\zeta+\zeta^2-3\zeta^3+6 \zeta^4+\\
&\quad\, (-\zeta^3+ 6\zeta^5)\log z)\, .
}[eqCeven]
Finally, since the RHS in~\eqref{cFshort}  depends only on one function, instead of evaluating the inversion integral, we can just compare the double discontinuity of $\CF^{\mathrm{short}}_{2244}$ and that of the supergravity answer.  First of all, we  check that we can recover the known supergravity result in the case of single-particle operators~\cite{Aprile:2017xsp,Caron-Huot:2018kta}
\eqna{
\dDisc\big[\CF_{2244}^{\mathrm{\lsp short,\,sp}}\big]\equiv\dDisc\big[\hat\CF_{2244}^{\mathrm{\lsp short,\,sp}}\big]=\dDisc \big[ \CH_{2244}^{\mathrm{free,\,sp}}(z,\zb) - \frac1c\,u^4 \Db_{4622}(z,\zb) \big]\, .
}[]
For the double-trace case, instead, we have two contributions
\twoseqn{
 \dDisc\big[\hat\CF_{2244}^{\mathrm{\lsp short,\,dt}}\big]&=\dDisc\left[\CH_{2244}^{\mathrm{free,\,dt}} - \frac1c\,u^4 \Db_{4622}\right] \,,
}[]
{
 \dDisc\big[\delta\CF_{2244}^{\mathrm{\lsp short,\,dt}}|_{\CC_{[0,2,0]\ell}}\big]&=\frac{1}{c} \dDisc\left[u^4 \Db_{4622}-2 u^2 \Db_{2422}\right]\,. 
}[][]
Summing them together we conclude that 
\eqna{
\CH^{\mathrm{sugra,dt}}_{2244}= 2+\frac{2}{v^2}+\frac{1}{c}\left( 1+\frac{1}{v^2}+\frac{6}{v}-2u^2 \Db_{2422}\right)\,. 
}[HE2244dt]
This result agrees with the one recently found in~\cite{Aprile:2024lwy}.
In hindsight, we could have expected a similar expression in virtue of the fact that, apart from normalization, $\CO_4^{\rm dt}$ is just the square of $\CO_2$. So that at tree level  the four-point function effectively factorizes as
\eqna{
\langle \CO_2(x_1)\CO_2(x_2) \CO_2(x_3)^2 \CO_2(x_4)^2\rangle \sim 4\langle \CO_2(x_3)\CO_2(x_4)\rangle \langle \CO_2(x_1)\CO_2(x_2) \CO_2(x_3) \CO_2(x_4)\rangle\, , 
}[]
and it is known that 
\eqna{
\CH_{2222}= 1+\frac{1}{v^2}+\frac{1}{c} \left(\frac{1}{v}-u^2 \Db_{2422} \right)\, .
}[]
For completeness in Appendix~\ref{app:OPEdecomp}, we report the decomposition in superconformal blocks for the supergravity results in both channels and for the correlators involving single-particle and double-trace operators.

\section{Solving mixing at twist six }\label{sec:twist6}
Armed with the supergravity results for $\CH_{2244}$ and the crossed channel, we can try to extract new OPE data for the long exchanged operators transforming respectively in the $[0,0,0]$ and $[0,2,0]$ representations.  Already in the known cases of correlators involving  single-particle operators only, extracting these data is incredibly difficult due to the fact that multiple operators with the same twist and spin can be exchanged in the OPE.  This leads to a ``mixing'' problem, which, for double-trace operators, has been solved at leading order in $1/c$ in~\cite{Alday:2017xua,Aprile:2017bgs,Aprile:2017xsp,Aprile:2018efk}. In this section, we will study how this problem manifests itself when correlators with external double-trace operators are considered,  whether higher-trace operators start to appear in the OPE and how to treat them.  In particular, we will focus on a subsector of long exchanged operator with twist six.  In fact, this is the minimal twist at which one can construct  triple-trace operators, like $[\CO_2, \CO_2, \CO_2]_\ell$.  To conclude we should mention that, even if one could extract the anomalous dimensions and OPE coefficients of these operators, that would be averages over degenerate states. In fact, differently from the double-trace case,  where, once we  fix twist and spin, there is only one primary, for primaries built out of three operators there is still some residual degeneracy for high enough spins. 
\subsection{Singlet R-symmetry channel}
The only long operators in the  $[0,0,0]$ representation with twist six that we can write down and that can potentially mix are schematically the double-trace operators
\eqna{
\CD_{i, \ell}&=\lbrace [\CO_2, \CO_2]_{1, \ell}, \, [\CO_3, \CO_3]_{0, \ell} \rbrace\, , \qquad i=1,2\, , \\
[\CO_p, \CO_q]_{n, \ell}&\equiv \CO_p \lsp \Box^n \partial_{\mu_1} \cdots \partial_{\mu_\ell} \CO_q\, ,
}[]
and the triple-trace one
\eqna{
\CT_{\ell}=[\CO_2, \CO_2, \CO_2]_{\ell}\, .
}[]
At large $c$, the double-trace operators get corrections to their anomalous dimensions and OPE coefficients with two $\CO_p$'s
\eqna{
\Delta_{\CD_{i, \ell}}&= 6+\ell+ \frac{ \gamma^{(1)}_i}{c}+\frac{ \gamma^{(2)}_i}{c^2}+\cdots\, , \\
\lambda_{pp\CD_{i, \ell}}^2&=(\lambda^{(0)}_{pp\CD_i})^2+\frac{(\lambda^{(1)}_{pp\CD_i})^2}{c}+\frac{(\lambda^{(2)}_{pp\CD_i})^2}{c^2}+\cdots \, .
}[]
We expect the triple-trace contributions to be suppressed in the large $c$ expansion with respect to the double-trace ones\footnote{One can intuitively see that from the fact that  the five-point function~\cite{Goncalves:2019znr} $\langle \CO_p\CO_p\CO_2\CO_2\CO_2\rangle \sim \frac{1}{\sqrt{c}}$ for $p=2,3$.  This intuition is also supported by the two-loop discussion in~\cite{Drummond:2022dxw}. Here the authors comment that in order to reproduce the $\log^2 u$ term in $\langle\CO_2 \CO_2 \CO_2 \CO_2  \rangle$ at order $c^{-2}$ is necessary to include the exchange of triple-trace operators.  The only way that these operators can appear at this order to this power of $\log u$ is if their OPE coefficient starts at $1/c$.}
\eqna{
\Delta_{\CT_{\ell}}&= 6+\ell+ \frac{ \Gamma^{(1)}}{c}+\frac{ \Gamma^{(2)}}{c^2}+\cdots\, , \\
\lambda_{pp\CT_{\ell}}^2&=\frac{(\lambda^{(1)}_{pp\CT})^2}{c}+\frac{(\lambda^{(2)}_{pp\CT})^2}{c^2}+\cdots \, .
}[]
Then we should plug this expansion in the OPE decomposition of $\CH$ in~\eqref{Hexp} projected to the singlet representation
\eqna{
\CH_{\{p_i\}}(z, \zb)^{\rm long}\Big|_{Y_{0,0}}&=\sum_{\Delta, \ell}a_{\{p_i\}}(\Delta, \ell;0,0) u^{\tau/2} \tilde{G}_{\Delta, \ell}^{(p_{12}, p_{34})}(z, \zb)=\CH^{(0)}+\sum_{\kappa=1}\frac{\CH^{(\kappa)}}{c^\kappa}\, , 
}[]
where we have used ${G}_{\Delta, \ell}^{(p_{12}, p_{34})}(z, \zb)=u^{\tau/2} \tilde{G}_{\Delta, \ell}^{(p_{12}, p_{34})}(z, \zb)$.  For $\CH_{2244}^{\rm dt}$, if we  define
\eqna{
a_{p4^{\rm dt}\CO} \equiv \lambda_{pp\CO}\lambda_{44\CO}^{\rm dt}\, ,  \qquad \quad \CO=\CD_i, \CT\, ,
}[]
then the contribution from twist six operators up to order $1/c$ reads
\eqna{
\CH^{(0)}&=u^3 a_{p4^{\rm dt}\CD_i}^{(0)} \tilde{G}_{6, \ell}(z, \zb) \, , \\
\CH^{(1)}&=\frac{u^3}{2}\log u \,   a_{p4^{\rm dt}\CD_i}^{(0)}\gamma_i^{(1)} \tilde{G}_{6, \ell}+u^3 \left(a_{p4^{\rm dt}\CD_i}^{(1)}+a_{p4^{\rm dt}\CT}^{(1)} +a_{p4^{\rm dt}\CD_i}^{(0)}\gamma_i^{(1)}   \partial_{\Delta}\right) \tilde{G}_{6, \ell}\, , \\
\CH^{(2)}&=\frac{u^3}{8}\log^2 u \,   a_{p4^{\rm dt}\CD_i}^{(0)}(\gamma_i^{(1)})^2 \tilde{G}_{6, \ell}+\frac{u^3}{2}\log u \Big(a_{p4^{\rm dt}\CD_i}^{(1)}\gamma_i^{(1)} +a_{p4^{\rm dt}\CD_i}^{(0)}\gamma_i^{(2)}+a_{p4^{\rm dt}\CT}^{(1)}\Gamma^{(1)} \\
&\quad\, +a_{p4^{\rm dt}\CD_i}^{(0)}(\gamma_i^{(1)})^2 \partial_\Delta \Big)\tilde{G}_{6, \ell}+u^3 \Big[a_{p4^{\rm dt}\CD_i}^{(2)}+a_{p4^{\rm dt}\CT}^{(2)} +\Big(a_{p4^{\rm dt}\CD_i}^{(1)}\gamma_i^{(1)}  +a_{p4^{\rm dt}\CD_i}^{(0)}\gamma_i^{(2)}   \\
&\quad\, +a_{p4^{\rm dt}\CT}^{(1)}\Gamma^{(1)}  \Big)\partial_{\Delta}+\frac{1}{2}a_{p4^{\rm dt}\CD_i}^{(0)}(\gamma_i^{(1)})^2 \partial_\Delta^2\Big] \tilde{G}_{6, \ell}\, , 
}[CH0&1]
where summations over $i$ and $\ell$ are understood and, since $\lambda_{22\CT}\sim 1/c$, we have assumed that $a_{p4^{\rm dt}\CT}$ has no disconnected part.

To solve the mixing problem at tree level, namely determining $a^{(0)},\, a^{(1)}, \,\gamma^{(1)}_i$ and $\Gamma^{(1)}$ we should consider a system of equations that take into account all correlators where twist-6 operators appear. In this example, this would require knowledge of all the four-point functions obtained by combining two $\CO_2$'s, $\CO_3$'s and $\CO_4^{\rm dt}$ up to order $c^{-2}$. Unfortunately this information is not available at the moment, but we can still make some interesting observations. First of all, we can try to say something about disconnected OPE coefficients. It is clear that $\langle \CO_4^{\rm dt}\CO_4^{\rm dt}\CT_\ell \rangle$ has no $c^{0}$ piece. Therefore, at this order, the only non-vanishing OPE coefficients are $\lambda_{44 \CD_i}^{(0), \text{dt}}$ with double-trace operators.  So if we combine the OPE coefficient in~\eqref{a02244}  with the result from expanding the disconnected four-point function $\CH_{4444}^{\rm dt}$, we obtain this system of equations at twist 6\footnote{The result for the disconnected contribution to the reduced correlator of four $\CO_4^{\rm dt}$ projected to the singlet R-symmetry channel reads
\eqna{
\CH^{\rm dt}_{4444}|_{Y_{0,0}}^{\rm disc}&=h(u,v)+\frac{1}{v^2}\lsp h\!\left(\frac{u}{v},\frac{1}{v} \right)\, ,\\
h(u,v)&=\frac{u^2}{30 v^2}+\frac{23 u^2}{20}-2 u v+\frac{8 u}{15 v}+\frac{34 u}{15}+\frac{9 v^2}{10}-\frac{14 v}{5}+\frac{12}{5 v}+\frac{13}{2} \, .
}[]}
\eqna{
\sum_{i=1}^2 \lambda_{22\CD_i}^{(0)}\lambda_{44\CD_i}^{\rm (0), dt} &=2\sum_{i=1}^2 (\lambda_{22\CD_i}^{(0)})^2\, , \\
\sum_{i=1}^2(\lambda_{44\CD_i}^{\rm (0), dt})^2 &=4\sum_{i=1}^2 (\lambda_{22\CD_i}^{(0)})^2\, .
}[a02244eqs]
A possible solution is simply $\lambda_{44\CD_i}^{\rm (0), dt}=2 \lambda_{22\CD_i}^{(0)}$ and the mixing for double-trace operators is luckily solved~\cite{Aprile:2017bgs}
\eqna{
\lambda_{22\CD_1}^{(0)}&=\sqrt{\frac{1+(-1)^\ell}{2} \frac{2^{\ell } (\ell +1) (\ell +2) (\ell +8) \Gamma (\ell +5)^2}{10 \Gamma (2 \ell +10)}}\, ,\\
\lambda_{22\CD_2}^{(0)}&=\sqrt{\frac{1+(-1)^\ell}{2} \frac{2^{\ell } (\ell +1) (\ell +7) (\ell +8) \Gamma (\ell +5)^2}{10 \Gamma (2 \ell +10)}}\, , \\
\gamma_1^{(1)}&=-\frac{120}{(\ell+1)(\ell+2)}\, , \qquad \gamma_2^{(1)}=-\frac{120}{(\ell+7)(\ell+8)}\,.
}[]
Another interesting observation we can draw is the difference between the correlator with $\CO_4^{\rm dt}$ and the one with only $\CO_2$'s at order $1/c$. In fact by  carefully removing the contributions from derivatives of the blocks, we can compute the average, $\langle\, \cdots \rangle=\sum_{\CO}$, of the corrections to the OPE coefficients and show how they differ.  For $\ell$ even
\threeseqn{
&\langle a^{(1)}_{22\CO} \rangle=\frac{1}{2} \partial_n \left(\langle a^{(0)}_{22\CO} \gamma^{(1)} \rangle \right)=\langle a^{(1), \text{free}}_{22\CO}\rangle+\langle a^{(1), \text{sugra}}_{22\CO}\rangle\, ,
}[]
{
&\langle a^{(1), \text{free}}_{22\CO}\rangle= (-2)^{\ell+1}\frac{(-1)^{n-1} \Gamma (n+1)^2 \Gamma (n+\ell +2)^2}{\Gamma (2 n+1) \Gamma (2 n+2 \ell +3)}\,,
}[]
{
&\begin{aligned}
&\langle a^{(1), \text{sugra}}_{22\CO}\rangle=- (-2)^{\ell+1}\frac{\Gamma (n+3)^2 \Gamma (n+\ell +2)^2}{(n+1) (n+2) \Gamma (2 n+1) \Gamma (2 n+2 \ell +3)} \times\\
&\times\! \left(\! (n-1) n \left(H_{2 n}-H_n+H_{2 (n+\ell +1)}-H_{n+\ell +1}\right)+\frac{1-(-1)^n -2 n^3-3 n^2+n}{(n+1)
   (n+2)} \right)\,,
\end{aligned}
}[a12222s][]
with $n=\tau/2$ and where we have used  $\langle a^{(1), \text{sugra}}\rangle$ to emphasize that it is the contribution   from the $\bar{D}$-function.  For the mixed correlator, we find for $\ell$ even
\twoseqn{
\langle a^{(1)}_{24^{\rm dt}\CO} \rangle &= \langle a^{(1), \text{free}}_{24^{\rm dt}\CO}\rangle+\langle a^{(1), \text{sugra}}_{24^{\rm dt}\CO}\rangle\, , \qquad \qquad \quad  \langle a^{(1), \text{sugra}}_{24^{\rm dt}\CO}\rangle=2  \langle a^{(1), \text{sugra}}_{22\CO}\rangle\, ,
}[]
{
\langle a^{(1), \text{free}}_{24^{\rm dt}\CO}\rangle&=- \frac{\left(\ell ^2+(2 n+3) \ell +2 \left(n-3 (-1)^{n+1}+1\right)\right) \Gamma (n+1)^2 \Gamma (n+\ell
   +2)^2}{(-2)^{-(\ell+1)}\Gamma (2 n+1) \Gamma (2 n+2 \ell +3)}\, ,
}[][]
which seems to violate the usual derivative relation~\cite{Heemskerk:2009pn}.

\subsection{$[0,2,0]$ R-symmetry channel}\label{subsec:020}
In the other channel, the interesting reduced correlators are
\eqna{
\CH_{4224}^{\rm dt}&=u\left( 2+\frac{1}{v^2}\right)+\frac{1}{c}\left( \frac{u(v+2)}{v}-2u^3\Db_{2422} \right)\, ,\\
\CH_{4224}^{\rm sp}&=\frac{u}{v^2}+\frac{1}{c}\left( \frac{2u}{v}-2u^3 v^2 \Db_{2642} \right)\, ,
}[020Rodd]
and the  long twist-6 operators we are looking for are in the [0,2,0] representation and they can be written schematically as
\eqna{
&\mathcal{D}_1=[\CO_4^{\rm sp}, \CO_2]_{0, \ell},\, \CT=[\CO_4^{\rm dt}, \CO_2]_{0, \ell}\sim[\CO_2, \CO_2, \CO_2]_{0, \ell}\ &&\quad \text{for }\ell\text{ odd} \, ,\\
&\mathcal{D}_1=[\CO_4^{\rm sp}, \CO_2]_{0, \ell},\, \CD_2=[\CO_3, \CO_3]_{0, \ell},\, \CT=[\CO_4^{\rm dt}, \CO_2]_{0, \ell} &&\quad \text{for }\ell\text{ even} \, .
}[]
As before, we denote by $\gamma_i^{(1)}$ and  $\Gamma^{(1)}$ the anomalous dimensions, at order $1/c$,  developed respectively by the operators $\CD_i$  and  $\CT$ and we use an expansion similar to~\eqref{CH0&1} to write down the unmixing equations. Since the mixing problem in the odd sector involves just two operators,  it turns out to be particular simple and it can actually be solved. To do that, it is enough to consider the four-point functions in~\eqref{020Rodd} and the fact that, by orthogonality, $\langle \CO_4^{\rm dt} \CO_2 \CO_2 \CO_4^{\rm sp}\rangle=0 $. Explicitly 
\eqna{
(\lambda_{42\CD_1}^{\rm dt (0)})^2+(\lambda_{42\CT}^{\rm dt (0)})^2&=-\frac{2^{\ell } (\ell +1) (\ell +3) (\ell +6) (\ell +8) \Gamma (\ell +6)^2}{15 (\ell +4) (\ell +5)
   \Gamma (2 \ell +9)}\, ,\\
  (\lambda_{42\CD_1}^{\rm sp (0)})^2+(\lambda_{42\CT}^{\rm sp (0)})^2&= -\frac{2^{\ell } (\ell +1) (\ell +8) \Gamma (\ell +6)^2}{15 \Gamma (2 \ell +9)}\,,\\
\lambda_{42\CD_1}^{\rm sp (0)} \lambda_{42\CD_1}^{\rm dt (0)}  +\lambda_{42\CT}^{\rm sp (0)} \lambda_{42\CT}^{\rm dt (0)}  &=0\, , \\
  (\lambda_{42\CD_1}^{\rm dt (0)})^2\gamma_1^{(1)}+(\lambda_{42\CT}^{\rm dt (0)})^2 \Gamma^{(1)}&=\frac{2^{\ell +1} (\ell +1) (\ell +3) (\ell +6) (\ell +8) \Gamma (\ell +6)^2}{(\ell +4)^2 (\ell +5)^2
   \Gamma (2 \ell +9)}\, ,\\
   (\lambda_{42\CD_1}^{\rm sp (0)})^2\gamma_1^{(1)}+(\lambda_{42\CT}^{\rm sp (0)})^2\Gamma^{(1)}&=  \frac{2^{\ell +1} (\ell +1) (\ell +8) \Gamma (\ell +6)^2}{(\ell +4) (\ell +5) \Gamma (2 \ell +9)}\, ,\\
   \lambda_{42\CD_1}^{\rm sp (0)} \lambda_{42\CD_1}^{\rm dt (0)} \gamma^{(1)} +\lambda_{42\CT}^{\rm sp (0)} \lambda_{42\CT}^{\rm dt (0)}\Gamma^{(1)}  &=0\, .
}[]
The solution for the anomalous dimensions is given by
\eqna{
\gamma^{(1)}_1=\Gamma^{(1)}=-\frac{30}{(\ell+4)(\ell+5)}\, ,
}[]
which for $\CD_1$ agrees with the results in~\cite{Aprile:2018efk}. For the OPE coefficients 
\eqna{
&\lambda_{42\CD_1}^{\rm sp (0)}=\sqrt{\frac{(\ell +4) (\ell +5)}{(\ell +3) (\ell +6)}} \lambda_{42\CT}^{\rm dt (0)}\, , \qquad \lambda_{42\CT}^{\rm sp (0)}=-\sqrt{\frac{(\ell +4) (\ell +5)}{(\ell +3) (\ell +6)}} \lambda_{42\CD_1}^{\rm dt (0)}\, ,\\
&\quad\,(\lambda_{42\CD_1}^{\rm dt (0)})^2+(\lambda_{42\CT}^{\rm dt (0)})^2=-\frac{2^{\ell } (\ell +1) (\ell +3) (\ell +6) (\ell +8) \Gamma (\ell +6)^2}{15 (\ell +4) (\ell +5)
   \Gamma (2 \ell +9)}\, .
}[]
Notice that a possible solution is $ \lambda_{42\CT}^{\rm sp (0)}=0= \lambda_{42\CD_1}^{\rm dt (0)}$, i.e.  the only operator exchanged in the OPE of $\CO_4^{\rm dt} \times \CO_2$ is the triple-trace one, while in $\CO_4^{\rm sp} \times \CO_2$ only the double-trace one. 

The solution of the mixing at even spins is more complicated, since we need to consider additional correlators, in particular $\langle\CO_3\CO_3\CO_3\CO_3\rangle$ and $\langle\CO_3\CO_3\CO_2\CO_4\rangle$ with $\CO_4$ both single-particle and double-trace. Since the latter is not known,\footnote{Potentially one could extract $\langle\CO_3\CO_3\CO_2\CO_4^{\rm dt}\rangle$ by taking the OPE of $\CO_2\times\CO_2$ inside the five-point function $\langle \CO_3\CO_3\CO_2\CO_2\CO_2\rangle$ in~\cite{Goncalves:2023oyx}.} we postpone this analysis to future work.

\section{One loop}\label{sec:oneloop}
So far we have shown that in one channel, the protected subsector of the correlator is the only contribution to dDisc and therefore it is enough to fix the tree-level supergravity result entirely.  It is legitimate to ask whether this information can be used to to infer one-loop OPE data as well.  This was indeed the case for the $\langle \CO_2\CO_2 \CO_2\CO_2\rangle$ discussed in~\cite{Alday:2017vkk}. In this example, the tree-level anomalous dimensions and OPE coefficients of double-trace operators,  upon careful unmixing,  are enough to compute $\log^2 v$ at order $c^{-2}$. Being the only contribution to dDisc, this piece of the crossed correlator completely fixes $c(\Delta, \ell)$ at one loop. For the correlators at hand the situation is more intricate.  For either extracting  OPE data through $c(\Delta, \ell)$ or reconstructing  the one-loop correlator directly using the dispersion relation in~\cite{Carmi:2019cub}, one would need
\eqna{
\text{Direct channel:}\, \CH_{2244}^{\rm (2), dt} &\,\leftrightarrow\quad \dDisc\left[ \frac{u^2}{v^3}\CH_{4224}^{\rm (2), dt}(v,u) \right]\, ,  \\
\text{Crossed channel:}\, \CH_{4224}^{\rm (2), dt} &\,\leftrightarrow\,  \left\lbrace \dDisc\left[ \frac{u^3}{v^2}\CH_{2244}^{\rm (2), dt}(v,u) \right], \,\dDisc\left[ \frac{u^3}{v^3}\CH_{2424}^{\rm (2), dt}(v,u) \right]\right\rbrace \, , 
}[]
where the superscript $(2)$ stands for order $c^{-2}$.  Similarly to the four-point function of $\CO_2$'s in~\cite{Alday:2017vkk}, for the direct channel the only contributions to dDisc comes from
\eqna{
\frac{1}{8}\log^2 v \sum v^{\tau /2}a^{(0)}\left(\gamma^{(1)}\right)^2\tilde{G}_{\tau+\ell, \ell}(v, u)\, , 
}[logvSqrd]
where the OPE data are the ones of long-operators exchanged in the [0,2,0] representation. In Sec.~\ref{subsec:020}, we have shown that, at least for twist six and odd spin,  among these operators we need to consider triple-trace ones, for which no unmixed information is available at the moment, thus making impossible to compute~\eqref{logvSqrd}.

Tackling the problem in the crossed channel is even harder. First of all we need to consider dDisc in two different channels and for the same reasons as above, little can be said about $\CH_{2424}^{\rm (2), dt}$. Second, given the fact that $p_{12}\neq 0$ and $p_{34}\neq 0$, lower powers of $\log v$ might contribute to dDisc, thus requiring knowledge of even more OPE data at previous order. Suppose nonetheless we can ignore these problems and we can assume that in $\CH_{2244}^{(1)}$ they appear only the same double-trace operators exchanged in $\CH_{2222}^{(1)}$ but with $a_{24^{\rm dt}\CO}^{(0)}=2a_{22\CO}^{(0)}$, as for the twist six in~\eqref{a02244eqs}.  Then we would know the part of $\dDisc\left[\CH_{2244}^{\rm (2), dt}(v,u)\right]$ coming from the $\log^2v$ generated by double-trace contributions. But an expression for this piece is already known since it is just $2\CH_{2222}^{(2)}|_{\log^2 v}$~\cite{Aprile:2017bgs, Alday:2017vkk,Caron-Huot:2018kta}. So we can deduce
\eqna{
\CH_{4224}^{\rm (2),dt }(u,v) \supset 2\CH_{2222}^{(2)}(v,u)\, ,
}[]
which is consistent with the results in~\cite{Aprile:2024lwy}. To make this statement more precise,  one could use the dispersion relation in~\cite{Carmi:2019cub}, which writes down the correlator as an integral of its dDisc over a kernel.  Unfortunately, this kernel for unequal external dimensions has not been worked out explicitly yet. 

\section{Future directions}\label{sec:futureDirections}
\begin{itemize}[leftmargin=*]
\item Higher loops: We have described how to construct four-point functions for double-trace operators up to order $c^{-1}$, only using information imposed by superconformal symmetry. It would be interesting to see how to compute higher terms in the large $c$ expansion. This is a challenging problem for several reasons, most of them being technical. As already mentioned in the text, there is mixing between intermediate long operators, which would require the knowledge of several correlators of single- and double-trace operators not known at the moment. In addition, computing the integrals appearing in the Lorentzian inversion formula for higher order is a challenging task, in particular since for correlators of non identical operators one needs to consider all the possible channels.
\item Triple trace operators: Disentangling the contribution of triple- and higher-trace operators is also an interesting future direction. In particular it would be interesting to see the structure of this class of operator for large spin, which is a completely unexplored arena. The challenges in this case are similar to the ones reported for higher loops, due to the fact that generically they mix with double-trace operators and solving the mixing, at generic values of the twist, is very complicated. 
\item An alternative approach to determining the anomalous dimensions for triple-trace operators could be to use the methods introduced in~\cite{Fardelli:2024heb}.  Within this framework,  corrections to the dimensions are computed  as energy eigenvalues of an Hamiltonian defined in AdS.  Although the analysis in~\cite{Fardelli:2024heb} is limited to non-supersymmetric theories, extending it to the case of interest should be straightforward --- particularly by considering states constructed from the $\CL_k$ superdescendants of the single-particle Kaluza-Klein half-BPS operators.
\item Composite operators in $\CN=2$ SCFTs: It would be interesting to extend  the analysis of double-trace operators in $\CN=4$ SYM to less supersymmetric $4d$ theories, particularly those with eight supercharges. Recent studies of holographic correlators in $AdS_5\times S^3$ involving supergluons --- the dimension-2 half-BPS scalar superpartners of spin-1 gluon --- and their Kaluza-Klein generalizations have yielded very interesting results, showing notable similarities to the four-point functions of single-particle operators in $\mathcal{N}=4$ SYM~\cite{Alday:2021odx,Alday:2021ajh,Alday:2022lkk,Alday:2023kfm,Behan:2024vwg,Huang:2023oxf,Bissi:2022wuh,Huang:2023ppy}. Investigating whether these analogies hold for multi-particle states in $\mathcal{N}=2$ SCFTs would be an interesting direction for future study.
\end{itemize}

\section*{Acknowledgements}
We thank Francesco Aprile, Stefano Giusto and Rodolfo Russo for correspondence and insightful discussions on an early version of this draft and on related topics. 
A.B.\ is partially supported by the Knut and Alice Wallenberg Foundation grant KAW 2021.0170, the Olle Engkvists Stiftelse grant 2180108 and  INFN Iniziativa Specifica ST\&FI. G.F.  is supported by the US Department of Energy Office of Science under Award Number DE-SC0015845, and is partially supported by the Simons Collaboration on the Non-perturbative Bootstrap.
\newpage
\appendix
\addtocontents{toc}{\protect\vskip1.5em}
\section{Higher-$p$ operators}\label{highertrace}
For any $p$, a single-particle operator is defined as the state which is orthonormal to any multi-particle state~\cite{Alday:2019nin,Aprile:2020uxk}.  \\
At $p=5$
\eqna{
\CO_5^{\rm sp}&=4N\sqrt{\frac{2(N^2+5)}{5(N-4)_9}}\left( t_5-\frac{5(N^2-2)}{N(N^2+5)}t_2t_3\right)\, ,\\
\CO_5^{\rm dt} &= \frac{4N}{\sqrt{3(N^2+5)(N-2)_5}}t_2t_3\,. 
}[]
where $(x)_n\equiv \frac{\Gamma(x+n)}{\Gamma(x)}$ stands for the Pochhammer symbol and we have defined $t_n\equiv {\rm tr}(y\cdot\phi)^n$
At $p=6$, also triple-trace operators start to appear and a possible basis of orthonormal operators is given by\footnote{Notice that it is not the unique choice.}
\eqna{
\CO_6^{\rm sp}&=\sqrt{\frac{(N^2-1)(N^4+15N^2+8)}{3N^3(N-5)_{11}}}\left( t^6-\frac{6(N^2-4)(N^2+5)}{N(N^4+15N^2+8)}t_2t_4\right.\\
&\quad\,\left.-\frac{3N^4-11N^2+80}{N(N^4+15N^2+8)t_3^2}+\frac{7(N^2-7)}{N^4+15N^2+8}t_2^3\right)\, ,\\
\CO_6^{\prime}&=\frac{1}{2N^3}\sqrt{\frac{N^6+25N^4+28N^2+18}{6(N^2-9)(N^2-1)(N^4+15N^2+8)}} \left( t_2^3\right.\\
&\quad\, \left.-\frac{6N(2N^4+19N^2+3)}{N^6+25N^4+28N^2+18}t_2t_4+\frac{8N(N^2-9)(5N^2+1)}{(N^2-4)(N^6+25N^4+28N^2+18)}t_3^2\right)\, ,\\
\CO_6^{\rm dt,(1)}&=-\frac{1}{2N^2}\sqrt{\frac{N^2+8}{N^6+25N^4+28N^2+18}}\left( t_2t_4-\frac{4(N^2-6)}{(N^2-4)(N^2+8)}t_3^2\right)\, ,\\
\CO_6^{\rm dt, (2)}&=\frac{1}{2N(N^2-4)\sqrt{N^2+8}}t_3^2\,.
}[]
Similarly, at $p=7$ and $p=8$ we checked that it is always possible to construct a basis in which, given $n$ double-trace building blocks, $n$ operators are only made of double traces.  More concretely at $p=7$, we can build two operators
\eqna{
\CO_7^{\rm dt, (1)}&=\alpha_1 t_2 t_5+\alpha_2 t_3t_4\, , \\
\CO_7^{\rm dt, (1)}&=\alpha_3  t_3t_4\, ,
}[]
and at $p=8$ one possible solution contains 
\eqna{
\CO_8^{\rm dt,(1)}&=\beta_1 t_2 t_6+\beta_2 t_3t_5+\beta_3 t_4^2\, , \\
\CO_8^{\rm dt,(2)}&=\beta_4 t_3t_5+\beta_5 t_4^2\, , \\
\CO_8^{\rm dt,(3)}&=\beta_6 t_4^2\, .
}[]
\section{General setting: conventions}
\label{app:conv}
In this appendix we are collecting the notations that we use throughout the text and some technical step needed to analyse the contribution of short and semi-short multiplet from the OPE decomposition of the four-point function of $\frac{1}{2}$-BPS operators.  Superconformal symmetry fixes the four-point function to have the following structure 
\eqn{
\langle \CO_{p_1}(x_1,y_1)\CO_{p_2}(x_2,y_2)\CO_{p_3}(x_3,y_3)\CO_{p_4}(x_4,y_4)\rangle = \CK_{\{p_i\}}(x_i,y_i) \;\CG_{\{p_i\}}(z,\zb;\alpha,\alphab)\,,
}[]
 with $p_4 \geq \max(p_1,p_2,p_3)$. The kinematical function $\CK_{\{p_i\}}(x_i,y_i)$ and the cross ratios built with the space time coordinates $u,v$ and with the polarizations  $\sigma, \tau$ are defined as
\threeseqn{
u = z \zb &= \frac{x_{12}^2 x_{34}^2}{x_{13}^2x_{24}^2}\,,\qquad
v = (1-z) (1-\zb) = \frac{x_{14}^2 x_{23}^2}{x_{13}^2x_{24}^2}\,,\qquad x_{ij}=x_i-x_j\,,
}[]{
\sigma = \alpha \alphab &= \frac{y_{13}\lsp y_{24}}{y_{12}\lsp  y_{34}}\,,\qquad
\tau = (\alpha-1)(\alphab-1) = \frac{y_{13}\lsp y_{23}}{y_{12}\lsp  y_{34}}\,,\qquad y_{ij} = y_i\cdot y_j\,,
}[]{
\CK_{\{p_i\}} &= \begin{aligned}[t]
&\frac{1}{(x_{12}^2)^{\frac12(p_1+p_2)}(x_{34}^2)^{\frac12(p_3+p_4)}}\left(\frac{x_{14}^2}{x_{13}^2}\right)^{\frac12 p_{34}}\left(\frac{x_{24}^2}{x_{14}^2}\right)^{\frac12 p_{12}}\\
& \times y_{12}^{\frac12(p_1+p_2+p_3-p_4)}\,y_{14}^{\frac12(p_{12}-p_{34})}\, y_{24}^{-\frac12(p_{12}+p_{34})} y_{34}^{p_3} \,,\qquad p_{ij}=p_i-p_j\,.
\end{aligned}
}[][]
The dynamical component of the four-point function is encoded in the function $\CG_{\{p_i\}}(z,\zb;\alpha,\alphab)$ and with the choice of prefactor $\CK_{\{p_i\}}$ as above, the function $\CG_{\{p_i\}}$ can be expanded in conformal and R-symmetry blocks as follows
\eqn{
\CG_{\{p_i\}}(z,\zb;\alpha,\alphab) = \sum_{n,m}\sum_{\Delta,\ell} C_{\{p_i\}}(\Delta,\ell;m,n)\,Y_{n,m}^{(p_{12},p_{34})}(\alpha,\alphab)\, g_{\Delta,\ell}^{(p_{12},p_{34})}(z,\zb)\,.
}[allCBexp]
This description encodes the fact that the intermediate operators transform in the $\SU(4)$ representations labeled by $n,m$ with Dynkin labels
\eqn{
[n-m,2m+\delta,n-m]\,,\qquad n=0,1,\ldots,p\,,\qquad m=0,1,\ldots,n\,.
}[Rrepr]
where we defined
\eqn{
\qquad \delta \equiv \max(|p_{12}|,|p_{34}|)\,,\qquad p \equiv \min\big(p_i,\tfrac12(p_i+p_j+p_k-p_l)\big)\,.
}
This is a consequence of the fact that 
\begin{equation}
[0,p_1,0] \times [0,p_2,0] \cap [0,p_3,0] \times [0,p_4,0]=\sum_{n=0}^p\sum_{m=0}^n [n-m,2m+\delta,n-m].
\end{equation}
The conformal blocks associated to the exchange of these conformal primaries are defined as
\twoseqn{
k_h^{(p_{12},\lsp p_{34})}(z) &= z^h \,{}_2F_1\mleft(h-\tfrac{p_{12}}2,h+\tfrac{p_{34}}2;\lsp 2h;\lsp z\mright)\,,
}[]{
g_{\Delta,\ell}^{(p_{12},\lsp p_{34})}(z,\zb) &= \frac{z \zb}{z-\zb}\left(-\frac12\right)^\ell\left(
k_{\frac{\Delta+\ell}2}^{(p_{12},\lsp p_{34})}(z)\; k_{\frac{\Delta-\ell-2}2}^{(p_{12},\lsp p_{34})}(\zb) - (z\leftrightarrow \zb)\right)\,,
}[][]
whereas the R-symmetry blocks are defined as
\eqn{
Y_{n,m}^{(p_{12},p_{34})}(\alpha,\alphab) = (-1)^{n-m} (\alpha\alphab)^{\frac12p_{34}}\, g_{-n-m-\delta,n-m}^{(-p_{12},\lsp -p_{34})}\mleft(\frac1\alpha,\frac1\alphab\mright)\,.
}
So far, the power of supersymmetry has not been exploited at its best. In particular the function $\CG_{\{p_i\}}$ satisfies the superconformal Ward identities. To cast them in a simpler form, let us introduce
\eqn{
\tilde\CG_{\{p_i\}}(z,\zb;\alpha,\alphab) \equiv (z\zb)^{\frac12p_{34}} \;\CG_{\{p_i\}}(z,\zb;\alpha,\alphab)\,.
}[]
Then we have
\twoseqn{
\tilde\CG_{\{p_i\}}\mleft(\frac1\alpha,\frac1\alphab;\alpha,\alphab\mright) &= k_{\{p_i\}}\,,
}[]{
\tilde\CG_{\{p_i\}}\mleft(z,\frac1\alphab;\alpha,\alphab\mright) &= k_{\{p_i\}}+ \left(\alpha-\frac1z\right)\lsp \hat{f}_{\{p_i\}}(z,\alpha)\,.
}[][kandF]
These constraints can be solved by introducing a function $\CH_{\{p_i\}}(z,\zb;\alpha,\alphab)$ as follows
\eqna{
\tilde\CG_{\{p_i\}}(z,\zb;\alpha,\alphab) &= k_{\{p_i\}} + \CG_{\{p_i\}}^{\hat{f}} + R_{\{p_i\}}\CH_{\{p_i\}}(z,\zb;\alpha,\alphab)\,,
}[]
with
\twoseqn{
\CG_{\{p_i\}}^{\hat{f}}(z,\zb;\alpha,\alphab) &= \begin{aligned}[t]
&\frac{(\alphab z - 1)(\alpha \zb -1)\big(\big(\alpha-\tfrac1z\big)\hat{f}_{\{p_i\}}(z,\alpha)+\big(\alphab-\tfrac1\zb\big)\hat{f}_{\{p_i\}}(\zb,\alphab)\big)}{(z-\zb)(\alpha-\alphab)} \lsp-\\
&\frac{(\alpha z - 1)(\alphab \zb -1)\big(\big(\alphab-\tfrac1z\big)\hat{f}_{\{p_i\}}(z,\alphab)+\big(\alpha-\tfrac1\zb\big)\hat{f}_{\{p_i\}}(\zb,\alpha)\big)}{(z-\zb)(\alpha-\alphab)}\,,
\end{aligned}
}[]{
R_{\{p_i\}}(z,\zb;\alpha,\alphab) &= (z\alpha -1 )(\zb\alpha -1 )(z\alphab -1 )(\zb\alphab -1 ) (z \zb)^{\frac12 p_{34}}\,.
}[][otherkandF]
The function $\hat{f}_{\{p_i\}}$ is expanded in one-variable conformal and R-symmetry blocks as follows
\eqn{
\hat{f}_{\{p_i\}}(z,\alpha) = \sum_{n=0}^{p-1}\sum_{\ell=0}^\infty b_{\{p_i\}}(n,\ell)\, y_n^{(p_{12},p_{34})}(\alpha)\,\mathfrak{g}_\ell^{(p_{12},p_{34})}(z)\,,
}[fexp]
with
\eqna{
y_n^{(p_{12},p_{34})}(\alpha)&=\alpha^{\frac12p_{34}} \lsp k_{\frac12 p_{34}-n}^{(-p_{12},-p_{34})}\mleft(\frac1\alpha\mright)\,,\\
\mathfrak{g}_\ell^{(p_{12},p_{34})}(z) &=z^{\frac12p_{34}}\lsp k_{\ell-\frac12p_{34}+1}^{(p_{12},\lsp p_{34})}(z)\,.
}[]
The function $\CH_{\{p_i\}}$ is expanded in two-variable superconformal blocks as follows
\eqn{
\CH_{\{p_i\}}(z,\zb;\alpha,\alphab) = \sum_{n=0}^{p-2}\sum_{m=0}^n\sum_{\Delta,\ell} a_{\{p_i\}}(\Delta,\ell; n,m)\,Y_{n,m}^{(p_{12},p_{34})}(\alpha,\alphab)\, G_{\Delta,\ell}^{(p_{12},p_{34})}(z,\zb)\,,
}[Hexp]
where the superconformal block is defined as
\eqn{
G_{\Delta,\ell}^{(p_{12},p_{34})}(z,\zb) = (z\zb)^{-2} g_{\Delta+4,\ell}^{(p_{12},p_{34})}(z,\zb)\,.
}[]
The function $\CH_{\{p_i\}}$ is not crossing symmetric but one needs to add a remainder term $\Delta_{\{p_i\}}$ defined as follows
\eqna{
\CH_{p_1p_2p_3p_4}(z,\zb;\alpha,\alphab) &= \frac{u^{\frac12(p_1+p_2)}}{v^{\frac12(p_2+p_3)}}\tau^{\frac12(p_1+p_2+p_3-p_4)-2}\times\\&\;\quad
\times\CH_{p_3p_2p_1p_4}\mleft(1-z,1-\zb;\frac{\alpha}{\alpha-1},\frac{\alphab}{\alphab-1}\mright) + \Delta_{p_1p_2p_3p_4}\,.
}[Hcrossing]
Such remainder term can be explicitly computed from, for example, the free theory values of $\CH_{\{p_i\}}$ and its crossed, or by the crossing of $\CG^{\hat{f}}_{\{p_i\}}$ which can be obtained from the chiral algebra.

The crossing $1\leftrightarrow2$ is instead much simpler because it does not have any extra piece
\eqn{
\CH_{p_1p_2p_3p_4}(z,\zb;\alpha,\alphab)= \frac{1}{v^{\frac12p_{34}+2}}
\CH_{p_2p_1p_3p_4}\mleft(\frac{z}{z-1},\frac{\zb}{\zb-1};1-\alpha,1-\alphab\mright)\,.
}[]

\section{Free theory results} \label{app:free}

\subsection{Channel $\langle 2244\rangle$}
\threeseqn{
 k_{2244}^{\mathrm{sp}} &= 1+\frac{10}{c}\,,}[]{
 \hat{f}_{2244}^{\lsp\mathrm{sp}} &= \frac1c\left(\frac{2z(4z-5)}{z-1} - \frac{6z^2\alpha}{z-1}\right)\,,}[]{
 \CH_{2244}^{\llsp \mathrm{sp}} &= \frac1{c\lsp v}\,.   }[][chfree2244sp]

\threeseqn{
k_{2244}^{\mathrm{dt}} &= 5+\frac{12}{c}\,,
}[]{
\hat{f}_{2244}^{\lsp\mathrm{dt}} &= \begin{aligned}[t]&
\frac{2 z (z^2-4 z+2)}{(z-1)^2} + \frac1c\frac{z (9 z^2-22 z+12)}{(z-1)^2}
\\&+ \alpha  \left(\frac{2 z^2 (z^2-2 z+2)}{(z-1)^2}+\frac1c\frac{z^2(z^2-8 z+8)}{(z-1)^2}\right)\,,
\end{aligned}
}[]{
\CH_{2244}^{\llsp \mathrm{dt}} &= 2+\frac{2}{v^2}+\frac1c\left(1+\frac6v+\frac1{v^2}\right)\,.
}[][chfree2244dt]

\subsection{Channel $\langle 4224\rangle$}

\threeseqn{
k_{4224}^{\mathrm{sp}} &= 1+\frac{10}{c}\,,
}[]{
\hat{f}_{4224}^{\lsp\mathrm{sp}} &= \frac{z(1-2z)}{(z-1)^2}+\frac1c\frac{2z(4z-5)}{z-1} + \alpha\left(\frac{z^2}{(z-1)^2}- \frac1c\frac{2z^2}{z-1}\right)\,,
}[]{
\CH_{4224}^{\llsp \mathrm{sp}} &= \frac{u}{v^2}+\frac1c\frac{2u}{v}\,.
}[][chfree4224sp]

\threeseqn{
k_{4224}^{\mathrm{dt}} &= 5+\frac{12}{c}\,,
}[]{
\hat{f}_{4224}^{\lsp\mathrm{dt}} &= \begin{aligned}[t]&
\frac{z (2 z^2-6 z+3)}{(z-1)^2}+\frac1c\frac{z (9 z-11)}{z-1}
\\&+\alpha  \left(\frac{z^2(2 z^2-4 z+3)}{(z-1)^2}+\frac1c\frac{z^2(z-3)}{z-1}\right)\,,
\end{aligned}
}[]{
\CH_{4224}^{\llsp \mathrm{dt}} &= 2 u+\frac{u}{v^2}+\frac1c\left(u+\frac{2u}v\right)\,.
}[][chfree4224dt]

\section{Comments on \texorpdfstring{$\lambda_{24^{\rm dt}\CC}$}{<24C>}}
\label{app:tensors}
To fix the OPE coefficients in Sec.~\ref{sec:NTE} unambiguously we can construct the differential operators appearing in the OPE. Consider the OPE between two generic operators
\eqna{
\CO_1 \times \CO_2=\sum_{\CO} \lambda_{12\CO} \frac{t^{\mathsf{OPE}}_{\CO}(y_1, y_2, \CD_y)}{(x_{12}^2)^{\frac{1}{2}(\Delta_1+\Delta_2+\Delta_{\CO})}} \CO(x_2, y)+O(x_{12}^2)\, ,
}[]
where $t^{\mathrm{OPE}}_{\CO}$ is a differential operator depending, in the case of interest, on the covariant derivative 
\eqna{
\CD_M=\left(2+y \cdot \frac{\partial}{\partial y}\right)\frac{\partial}{\partial y^M}-\frac{1}{2}y_M\frac{\partial^2}{\partial y \cdot \partial y}\, .
}[]
Let us suppress the spacetime dependence and let us focus on the R-symmetry part of the OPE 
\eqna{
\CO_2 \times \CO_2 |_{[0,2,0]}&=\lambda_{2 2 [0,2,0]}\,  t^{\mathsf{OPE}}_2(y_1, y_2, \CD_y)\lsp \CO_{[0,2,0]}(y)\, , \\
\CO_2 \times \CO_4^{\mathrm{dt}} |_{[0,2,0]}&=\lambda_{2 4^{\mathrm{dt}} [0,2,0]}\,  t^{\mathsf{OPE}}_4(y_1, y_2, \CD_y)\lsp \CO_{[0,2,0]}(y)\, ,
}[]
such that
\eqna{
t^{\mathsf{OPE}}_2(y_1, y_2, \CD_y)\langle  \CO_{[0,2,0]}(y) \CO_{[0,2,0]}(y_3) \rangle &= y_{12} \lsp y_{13} \lsp y_{23}\, , \\
t^{\mathsf{OPE}}_4(y_1, y_2, \CD_y)\langle  \CO_{[0,2,0]}(y) \CO_{[0,2,0]}(y_3) \rangle &=( y_{12})^2 \lsp ( y_{23})^2\, .
}[]
This requirement completely fixes
\eqna{
t^{\mathsf{OPE}}_2(y_1, y_2, \CD_y)&=\frac{1}{12} y_{12} \lsp y_1 \cdot \CD_y \lsp y_2\cdot \CD_y\, ,\\
t^{\mathsf{OPE}}_4(y_1, y_2, \CD_y)&=\frac{1}{12} y_{12}^2 \left(y_2 \cdot \CD_y \right)^2\, .
}[]
Then we need to see how these conventions are reflected in the four-point function
\eqna{
\langle \CO_2\CO_2\CO_2 \CO_2 \rangle |_{[0,2,0]}&=\lambda_{22\CO}^2 t^{\mathsf{OPE}}_2(y_1,y_2, y) \langle \CO(y)\CO(y^\prime) \rangle t^{\mathsf{OPE}}_2(y_3,y_3, y^\prime)\\
&=\lambda_{22\CO}^2 \lsp \frac{1}{6} y_{12}^2 y_{34}^4 (-1+3\sigma+3\tau) \, ,\\
\langle \CO_2\CO_2\CO_2 \CO_4^{\mathrm{dt}} \rangle |_{[0,2,0]}&=\lambda_{22\CO}\lambda_{\CO 24^{\mathrm{dt}}} t^{\mathsf{OPE}}_2(y_1,y_2, y) \langle \CO(y)\CO(y^\prime) \rangle t^{\mathsf{OPE}}_4(y_3,y_3, y^\prime)\\
&= \lambda_{22\CO}\lambda_{\CO 24^{\mathrm{dt}}}\lsp y_{12}\lsp y_{14}\lsp y_{24} \lsp y_{34}^2\, , \\
\langle \CO_2\CO_4^{\mathrm{dt}}\CO_2 \CO_4^{\mathrm{dt}} \rangle |_{[0,2,0]}&=\lambda_{24^{\mathrm{dt}}\CO}^2 t^{\mathsf{OPE}}_4(y_1,y_2, y) \langle \CO(y)\CO(y^\prime) \rangle t^{\mathsf{OPE}}_4(y_3,y_3, y^\prime)\\
&= \lambda_{24^{\mathrm{dt}}\CO}^2\lsp y_{12}^2 y_{24}^2 y_{34}^2\, .
}[]
For the second and third case, we are using the same normalization, while in the first case if we want to use the results from~\cite{Dolan:2004iy} for $\lambda_{22\CC}$, we need to multiply by 6 and indeed we find
\twoseqn{
\lambda_{222}&= 6 \cdot \frac{1}{ 3 c }= \frac{2}{c} \, , }[]
{ \lambda_{22\CC}^2&= 6 \cdot \frac{1}{24}\lsp 2^{\ell +1} \left(1+(-1)^{\ell }\right)\frac{ (\ell +2)! (\ell +3)! }{(2 \ell +5)!}\left((\ell +3) (\ell
   +4)+\frac{1}{c}\right)\, .
}[l22CSqr][]
Then, given \eqref{lambdaC24},
\eqna{
 \lambda_{\CC 24^{\mathrm{dt}}}^2=\left(\frac{(-2)^{\ell+2} b_{\ell+2}}{\sqrt{\lambda_{22\CC}^2}}\right)^2 \xrightarrow{\text{ large } c\,} 2^{\ell } \left((-1)^{\ell }+1\right)\frac{(\ell +2)! (\ell +4)!}{c \lsp (2 \ell +6)!}.
}[]

As a further consistency check we can  compute explicitly the OPE coefficients involving $\CC_{[0,2,0], \ell}$ with spin zero. From~\cite{Arutyunov:2000ku}, we know that $\CC_{[0,2,0],0}$ is a double-trace operators constructed from $\CO_2$ 
\eqna{
\CC_{[0,2,0], 0}(x, y)&=\frac{3}{\sqrt{10 c (12c+1)}}y_M y_N \delta_{PQ} \lsp \CO_2^{MP}(x) \CO_2^{NQ}(x)\, ,\\
\CO_2^{M_1 M_2}(x)&=\mathrm{tr}(T^{I_1}T^{I_2}) \left(\varphi^{(M_1}_{I_1}(x)\varphi^{M_2)}_{I_2}(x) -\frac{1}{6}\delta^{M_1M_2} \varphi^{P}_{I_1}(x)\varphi_{P, I_2}(x)\right)
}[]
where $M,N,P, Q$ are fundamental SO(6) index, $T^I$ are SU($N$) matrices and $\varphi^M_I$ is the fundamental scalar of $\CN=4$ SYM. 
The normalization of the operator is chosen in such a way that its two-point function is unit-normalized
\eqna{
\langle \CC_{[0,2,0], 0}(x_1, y_1)\CC_{[0,2,0], 0}(x_2, y_2)\rangle=\frac{y_{12}^2}{(x_{12}^2)^4}\, .
}[]
With this definition we can compute the interesting three-point functions
\threeseqn{
\langle \CO_2\CO_2\CC_{[0,2,0],0}\rangle&=\sqrt{\frac{12c+1}{10c}} \frac{y_{12}\lsp y_{13}\lsp y_{23}}{(x_{12}^2 x_{13}^2)^2}\, ,
}[l22C]
{\langle \CC_{[0,2,0],0} \CO_2\CO_4^{\mathrm{dt}}\rangle&=4\sqrt{\frac{2c+1}{5c(12c+1)}} \frac{y_{13}^2 \lsp y_{23}^2}{x_{12}^2 (x_{13}^2)^3 x_{23}^2}\, ,
}[lC24dt]
{\langle \CC_{[0,2,0],0} \CO_2\CO_4^{\mathrm{sp}}\rangle&=0\,.
}[][]
Notice that~\eqref{l22C} coincides exactly with the square root of~\eqref{l22CSqr},  \eqref{lC24dt} is the square of~\eqref{lC24dtSqr} and the third one is zero as expected.

\section{Conformal block expansion of \texorpdfstring{$\boldsymbol{\hat{f}}$}{hatf}} \label{app:fhatexp}

Notice that, in general, when $p=2$ the function $\hat{f}_{\{p_i\}}$ is linear in $\alpha$ and the Jacobi polynomial for $n=0$ is just $1$ while that for $n=1$ is
\eqn{
y_1^{-p_{34},p_{34}}(\alpha) = \alpha - \frac{1}{2-p_{34}}\,.
}[]
Recall that in our choice of conventions $p_4$ is always the largest and therefore ${p_{34}\leq 0}$.

\subsection{Channel $\langle 2244\rangle$}

For this channel we have the expansion
\eqn{
\hat{f}_{2244}(z,\alpha) = \sum_{\ell=0,2,4,\ldots} b_{2244}(0,\ell)\,k_{\ell+1}^{(0,0)}(z) + \left(\alpha-\frac12\right)\sum_{\ell=1,3,5,\ldots}b_{2244}(1,\ell)\,k_{\ell+1}^{(0,0)}(z)\,.
}[]
\twoseqn{
b_{2244}^{\mathrm{sp}}(0,\ell) &= \frac{10}{c}\frac{(\ell!)^2}{(2\ell)!}
\,,
}[]{
b_{2244}^{\mathrm{sp}}(1,\ell) &= \frac{12}{c}\frac{(\ell!)^2}{(2\ell)!}\,.
}[][b2244sp]

\twoseqn{
b_{2244}^{\mathrm{dt}}(0,\ell) &= \frac{(\ell!)^2}{(2\ell)!}\left(2(1-\ell)(2+\ell) +\frac1c(3-\ell)(\ell+4)\right)
\,,
}[]{
b_{2244}^{\mathrm{dt}}(1,\ell) &= \frac{(\ell!)^2}{(2\ell)!}\left(4\ell(\ell+1)+\frac2c(\ell^2+\ell+6)\right)\,.
}[][b2244dt]

\subsection{Channel $\langle 4224\rangle$}
\eqn{
\hat{f}_{4224}(z,\alpha) = \sum_{\ell=0}^\infty b_{4224}(0,\ell)\,k_{\ell+2}^{(2,-2)}(z) + \left(\alpha-\frac14\right)\sum_{\ell=0}^\infty b_{4224}(1,\ell)\,k_{\ell+2}^{(2,-2)}(z)\,.
}[]
\twoseqn{
b_{4224}^{\mathrm{sp}}(0,\ell) &= \frac{(\ell!)^2(\ell+2)}{(2\ell+1)!}\left(
\frac{5 (\ell ^2+3 \ell +6 (-1)^{\ell }+2)}{8 c}-\frac{(\ell -1) (\ell +1) (\ell +2) (\ell +4)}{16}
\right)
\,,
}[]{
b_{4224}^{\mathrm{sp}}(1,\ell) &= \frac{(\ell!)^2(\ell+2)}{(2\ell+1)!}\left(
\frac{(\ell ^2+3 \ell -2 (-1)^{\ell }+2)}{2 c}+\frac{\ell  (\ell +1) (\ell +3) (\ell +2)}{12}
\right)\,,
}[][]

\twoseqn{
b_{4224}^{\mathrm{dt}}(0,\ell) &= \frac{(\ell!)^2(\ell+2)}{(2\ell+1)!}\begin{aligned}[t]
&\left(
\frac{-(-1)^{\ell } (\ell ^2-3\ell+34) + 5 \ell ^2 +15 \ell +10}{8 c}\right.\\
&\quad\left.-\frac{1}{16} (\ell -1) (\ell +4) (\ell ^2+3 \ell +4 (-1)^{\ell }+2)
\right)\,,
\end{aligned}
}[]{
b_{4224}^{\mathrm{dt}}(1,\ell) &= \frac{(\ell!)^2(\ell+2)}{(2\ell+1)!}\begin{aligned}[t]
&\left(
\frac{1}{12} \ell  (\ell +3) \left(\ell ^2+3 \ell -12 (-1)^{\ell }+2\right)\right.\\&\quad\left.-\frac{((-1)^{\ell }-1) (\ell +1) (\ell +2)}{2 c}
\right)\,.
\end{aligned}
}[][]
\section{Resummation to $\CH^{\rm short}$}\label{app:Hshort}
\subsection{Channel $\langle 2244\rangle$}
\eqna{
\CH_{2244}^{\mathrm{short,\,sp}}&=\frac{1}{c}\Bigg(	\frac{30(z\zb-z-\zb+2)}{z \zb (1-z)(1-\zb)} +\frac{12(5\zb(\zb-2)-z(2\zb^2+5\zb-10))}{z^2 \zb(1-\zb)(z-\zb)}\log(1-z)\\
&\quad\, +\frac{60}{(z\zb)^2} \log(1-z)\log(1-\zb)+z\leftrightarrow\zb\Bigg)\, , }[]

\eqna{
\CH_{2244}^{\mathrm{short,\,dt}}&=\frac{4 \left(2 z \zb^4-z \zb^3+4 z \zb^2-18 z \zb+12 z-3 \zb^4+18
   \zb^2-12 \zb\right)}{z^2\zb(1-\zb)^2(z-\zb)}\log(1-z)\\
   &\quad\, +\frac{12 \left(z^3-6 z+4\right)}{(1-z)^2 {\zb} (z-{\zb})}+\frac{24}{(z\zb)^2}\log(1-z)\log(1-\zb)\\
   &\quad\, +\frac{2}{c}\Bigg[\frac{3 (z-2) \left(z^2+12 z-12\right)}{(1 - z)^2 \zb (z - \zb)}+	\frac{36}{(z\zb)^2} \log(1-z)\log(1-\zb)\\
   &\quad\, +\left(\frac{2 \zb \left(\zb^2+4 \zb-4\right)}{z (\zb-1)^2 (z-\zb)}+\frac{3 \left(\zb^3+10 \zb^2-36 \zb+24\right)}{z^2 (\zb-1)^2 \zb}\right)\log(1-z)\Bigg]+z\leftrightarrow\zb
}[]
\subsection{Channel $\langle 4224\rangle$}
\eqna{
&\hat{\CH}^{\rm short, sp}_{4224}=\frac{4}{u^2} \Bigg \lbrace \frac{h_1(z, \zb)}{6 v^2}+\frac{h_2(z, \zb)}{(z-\zb)}\log(1-z)-\frac{h_2(\zb, z)}{(z-\zb)}\log(1-\zb) \\
&\quad\, +60 \frac{v^2}{u}\log(1-z)\log(1-\zb)\Bigg\rbrace +\frac{8}{c u^2}\Bigg\lbrace \frac{h_3(z, \zb)}{6v} +\frac{h_4(z, \zb)}{z-\zb}\log(1-z) \\
&\quad\,-\frac{h_4(\zb, z)}{z-\zb}\log(1-\zb)+\frac{300 v^2}{u}\log(1-z)\log(1-\zb)\Bigg\rbrace\, ,
}[CHshort4224sp]
where we have defined
\eqna{
h_1(z, \zb)&=(626 z^3 \zb^3-2998 z^3 \zb^2+2553 z^3 \zb-750 z^3+3344 z^2 \zb^2-5460 z^2
   \zb\\
  & \quad\, +1560 z^2+2205 z \zb-1260 z+180)+z\leftrightarrow\zb\, ,\\
  h_2(z, \zb)&=\frac{(1-z)^2}{z(1-\zb)^2}(5\zb(25 \zb^3-52 \zb^2+42 \zb-12)+z(12 \zb^4-125 \zb^3+260 \zb^2-210 \zb+60))\, , \\
  h_3(z, \zb)&=(3 z^3 \left(52 \zb^3-365 \zb^2+411 \zb-150\right)+z^2 \left(3079 \zb^2-8250
   \zb+3300\right)\\
   &\quad\,  +1125 z (5 \zb-4)+900)+z\leftrightarrow \zb\, , \\
   h_4(z, \zb)&=\frac{(1-z)^2}{1-\zb}\left(\!-18 \zb^4-75 \zb^3+550 \zb^2-750 \zb+300+\frac{25 \zb}{z} \left(3
   \zb^3-22 \zb^2+30 \zb-12\right)\!\right)\, .
}[]

\eqna{
&\hat{\CH}^{\rm short, dt}_{4224}=\frac{4}{u^2}\Bigg\lbrace \frac{h^\prime_1(z, \zb)}{6v^2}+\frac{h^\prime_2(z, \zb)}{z-\zb}\log(1-z)-\frac{h^\prime_2(\zb, z)}{z-\zb}\log(1-\zb)\\
&\quad\, +\frac{180 v^2}{u}\log(1-z)\log(1-\zb)\Bigg\rbrace+\frac{4}{c u^2}\Bigg\lbrace \frac{h_3^\prime(z, \zb)}{6 v} +\frac{h^\prime_4(z, \zb)}{z-\zb}\log(1-z)\\
&\quad\,-\frac{h^\prime_4(\zb, z)}{z-\zb}\log(1-\zb)+\frac{660 v^2}{u}\log(1-z)\log(1-\zb)\Bigg\rbrace\, ,
}[CHshort4224dt]
with
\eqna{
h^\prime_1(z, \zb)&=(2 z^5 (\zb-1)^2 \left(28 \zb^2-57 \zb+30\right)-4 z^4 \left(21 \zb^4-9
   \zb^3-82 \zb^2+93 \zb-30\right)\\
   &\quad\, +z^3 \left(2042 \zb^3-9366 \zb^2+7779
   \zb-2250\right)+12 z^2 \left(841 \zb^2-1365 \zb+390\right)\\
   &\quad\, +945 z (7
   \zb-4)+540)+z\leftrightarrow \zb\, ,\\
 h^\prime_2(z, \zb)&  =\frac{(1-z)^2}{z(1-\zb)^2}(z \left(4 \zb^6+2 \zb^5+36 \zb^4-375 \zb^3+780 \zb^2-630
   \zb+180\right)\\
   &\quad \, -5 \zb \left(2 \zb^5+4 \zb^4-75 \zb^3+156 \zb^2-126
   \zb+36\right))\, ,\\
   h^\prime_3(z,\zb)&=(z^4 \left(28 \zb^3-85 \zb^2+87 \zb-30\right)+z^3 \left(298 \zb^3-2313
   \zb^2+2679 \zb-990\right)\\
   &\quad\, +z^2 \left(6757 \zb^2-18150 \zb+7260\right)+2475 z
   (5 \zb-4)+1980)+z\leftrightarrow \zb\, ,\\
h^\prime_4(z, \zb)&  =\frac{(1-z)^2}{z(1-\zb)}(5 \zb \left(\zb^4+33 \zb^3-242 \zb^2+330 \zb-132\right)\\
&\quad\, -z \left(2
   \zb^5+39 \zb^4+165 \zb^3-1210 \zb^2+1650 \zb-660\right))\, . 
}[]
\section{Extracting OPE data} \label{app:OPEdecomp}
In this appendix we collect the results for the OPE decomposition of the four-point functions of $\CO_2$'s and $\CO_4$'s in the various channels for both single-particle and double-trace operators.
\subsection{Channel $\langle 2244\rangle$}
Let us start from the double-trace case \eqref{HE2244dt}, if we compare it to the result for the all $\CO_2$'s case
\eqna{
\CH_{2222}= 1+\frac{1}{v^2}+\frac{1}{c} \left(\frac{1}{v}-u^2 \Db_{2422} \right)\, ,
}[]
we notice that
\eqna{
\langle a^{(0)}\rangle_{2244}^{\mathrm{dt}}&=2 \langle a^{(0)}\rangle_{2222}\\
&=\frac{2^{\ell +1} \left((-1)^{\ell }+1\right) (\ell +1) \Gamma \mleft(\frac{\tau }{2}+1\mright)^2 (\tau
   +\ell +2) \Gamma \mleft(\ell +\frac{\tau }{2}+2\mright)^2}{\Gamma (\tau +1) \Gamma (2 \ell +\tau +3)}\, ,
}[a02244]
starting from twist 0 and where we should remember that these twist 0 and 2 are the ones necessary to cancel the contributions in $k$ and $\CG^{\hat{f}}$. 

From the $\log u$ part of~\eqref{HE2244dt}, we can obtain
\eqna{
\langle a^{(0)}\gamma^{(1)}\rangle^{\mathrm{dt}}_{2244}= -\frac{(\tau -2) \tau  (\tau +2) (\tau +4)}{16 (\ell +1) (\tau +\ell +2)} \langle a^{(0)}\rangle^{\mathrm{dt}}_{2244}\, ,
}[]
where the first non-zero twist contribution correspond to $\tau=4$.

Now for the single-particle correlator
\eqna{
\CH^{\mathrm{sp}}_{2244}=\frac{1}{c}\left( \frac{6}{v}-u^4\Db_{4622}\right)\, ,
}[]
where notice that there is no disconnected term. For the anomalous part instead,  we start at twist 8
\eqna{
\langle a^{(0)} \gamma^{(1)}\rangle_{2244}^{\mathrm{sp}}=\frac{ \left((-1)^{\ell }+1\right) }{2}\frac{(\tau -6) (\tau -4) (\tau -2) 2^{\ell -7}\Gamma \mleft(\frac{\tau
   }{2}+1\mright) \Gamma \mleft(\frac{\tau }{2}+5\mright) \Gamma \mleft(\ell +\frac{\tau }{2}+2\mright)^2}{3
   \Gamma (\tau ) \Gamma (2 \ell +\tau +3)}\, .
}[]
\subsection{Channel $\langle 4224\rangle$}
Let us start again from the double-trace contribution first 
\eqna{
\CH^{\mathrm{dt}}_{4224}=2 u+\frac{u}{v^2}+\frac{1}{c}\left(u+2\frac{u}{v}-2 u^3 \Db_{2422}\right)\, .
}[]
The disconnected contribution now starts at twist 2
\eqna{
\langle a^{(0)} \rangle^{\mathrm{dt}}_{4224}&=\frac{2^{\ell -2} (\ell +1) \Gamma \left(\frac{\tau }{2}+2\right) \Gamma \left(\frac{\tau }{2}\right)
   (\tau +\ell +2) \Gamma \left(\ell +\frac{\tau }{2}+1\right) \Gamma \left(\ell +\frac{\tau
   }{2}+3\right)}{3 \Gamma (\tau +1) \Gamma (2 \ell +\tau +3)}\times
  \\
  &\quad \, \times \left(24+(-1)^{\ell } \frac{1}{16} \tau  (\tau +2)  (\tau +2 \ell +2) (\tau +2 \ell +4) \right)\\
  &=\langle a^{(0)}\rangle_{4224}^{\mathrm{sp}}\left( 1+\frac{384 (-1)^{\ell }}{\tau  (\tau +2) (\tau +2 \ell +2) (\tau +2 \ell +4)} \right)
  }[]
For the anomalous dimension we get, starting from $\tau=6$, 
\eqna{
&\langle a^{(0)}\gamma^{(1)}\rangle_{4224}^{\mathrm{dt}}=-\frac{(\tau -4) (\tau -2) (\tau +4) (\tau +6)}{4 (\tau +2 \ell +2) (\tau +2 \ell +4)} \langle a^{(0)} \rangle_{4224}^{\mathrm{dt}}\times\\ &\quad\, \Big \lbrace \frac{1-(-1)^\ell}{2} \frac{\ell ^2+(\tau +3)
   \ell+\frac{(\tau -2) \left(\tau ^3+16 \tau ^2+148 \tau +480\right)}{10 \tau  (\tau +2)}}{\ell ^2+(\tau +3) \ell+\frac{1}{4} (\tau +2) (\tau +4)-\frac{96}{\tau  (\tau +2)}}+
  \\
  &\quad \, +\frac{1+(-1)^\ell}{2} \frac{1}{(1+\ell)(\tau +\ell +2)(\ell ^2+(\tau +3) \ell+\frac{1}{4} (\tau +2) (\tau +4)+\frac{96}{\tau  (\tau +2)})}\times\\
  &\quad \, \Big( \ell^4+2 (\tau +3) \ell ^3+\ell^2\left(\frac{1}{10} (\tau  (11 \tau +82)+202)+\frac{96}{\tau  (\tau +2)}\right)+\\
  &\quad \, \frac{(\tau +3) (\tau +12) (\tau  (\tau  (\tau +12)+12)+80) \ell }{10 \tau  (\tau +2)}+\frac{(\tau +2) (\tau +8) (\tau  (\tau +4)+60)}{10 \tau }\\
  &\quad\, +\frac{(\tau +3) (\tau +12) (\tau  (\tau  (\tau
   +12)+12)+80) \ell }{10 \tau  (\tau +2)} \Big)\Big\rbrace\, .
}[]
Now passing to the single-particle case, starting at twist 2 we have
\eqna{
\langle a^{(0)} \rangle_{4224}^{\mathrm{sp}}=\frac{(-1)^{\ell } 2^{\ell -2} (\ell +1) \Gamma \left(\frac{\tau }{2}+2\right)^2 (\tau +\ell +2) \Gamma
   \left(\ell +\frac{\tau }{2}+3\right)^2}{3 \Gamma (\tau +1) \Gamma (2 \ell +\tau +3)}\, ,
}[]
and for the anomalous dimensions starting at twist 6
\eqna{
\langle a^{(0)}\gamma^{(1)} \rangle_{4224}^{\mathrm{sp}}&=\frac{(\tau -4) (\tau -2) (\tau +4) (\tau +6)}{4 (\tau +2 \ell +2) (\tau +2 \ell +4)}\langle a^{(0)} \rangle_{4224}^{\mathrm{sp}} \Big\lbrace \frac{1-(-1)^\ell}{2}\\
&\quad\,  +\frac{1+(-1)^\ell}{2}\frac{(\tau +2)^2+2 \ell ^2+2 (\tau +3) \ell }{2 (\ell +1) (\tau +\ell +2)} \Big\rbrace
}[]

\section{Double discontinuity and inversion formula} \label{app:inversion}
Suppose for simplicity that the in the reduced function $\CH_{\{p_i\}}$ only one representation is exchanged. Then we introduce the function $c_{\{p_i\}}(\Delta,\ell)$ as the OPE coefficient density obtained by applying the Lorentzian inversion integral to $\CH_{\{p_i\}}$. More precisely we have~\cite{Caron-Huot:2017vep}
\eqn{
c_{\{p_i\}}(\Delta,\ell) = \int_0^1\di z\di\zb\, \mu^{(p_{12},p_{34})}_{\Delta,\ell}(z,\zb)\,\dDisc\lsp\mleft[\CH_{\{p_i\}}(z,\zb)\mright] + (-1)^\ell(p_1\leftrightarrow p_2)\,.
}[invFor]
The measure $\mu_{\Delta,\ell}$ is given by
\eqn{
\mu^{(p_{12},p_{34})}_{\Delta,\ell}(z,\zb) = \frac{1}{4}\kappa^{(p_{12},p_{34})}_{\frac{\Delta+\ell+4}2}\frac{(z-\zb)^2}{(z\zb)^4}\,g^{(-p_{12},-p_{34})}_{\ell+3,\Delta-3}(z,\zb)\,,
}[]
with
\eqn{
\kappa_h^{(r,s)} = \frac{\Gamma\mleft(h+\frac12r\mright)\Gamma\mleft(h-\frac12r\mright)\Gamma\mleft(h+\frac12s\mright)\Gamma\mleft(h-\frac12s\mright)}{2\pi^2\, \Gamma(2h-1)\Gamma(2h)}\,.
}[]
The double-discontinuity is defined as the difference between the Euclidean correlators and its two possible analytic continuations around $\zb=1$
\eqna{
\dDisc[\CG(z, \zb)]&=\cos\left( \pi \alpha\right)\CG(z, \zb)-\frac{1}{2}\left( e^{i\pi \alpha}\CG^\circlearrowleft(z, \zb)+e^{-i\pi \alpha}\CG^\circlearrowright(z, \zb)\right)\, ,  \\
\alpha&\equiv \frac{p_{34}-p_{12}}{2}\,.
}[]
The poles and residues of $c_{\{p_i\}}(\Delta,\ell)$ give the spectrum of $\CH_{\{p_i\}}$
\eqn{
\lim_{\Delta\to\Delta_k} c_{\{p_i\}}(\Delta,\ell) = \frac{a_{\{p_i\}}(\Delta_k,\ell)}{\Delta_k-\Delta}\,.
}[]

\Bibliography[refs.bib]
\end{document}